\documentclass[12pt]{article}
\usepackage[english]{babel}
\usepackage{amsmath}
\usepackage{cite}
\usepackage{amsfonts}
\usepackage{amssymb}
\begin{document}

\title{\hfill{\normalsize{}hep-th/0608005}\\[5mm]
{\bf{}Gauge invariant Lagrangian formulation of higher spin massive
bosonic field theory in AdS space}}

\author{\sc I.L. Buchbinder${}^{a}$\thanks{joseph@tspu.edu.ru},
V.A. Krykhtin${}^b$\thanks{krykhtin@mph.phtd.tpu.edu.ru}, P.M.
Lavrov$^c$\thanks{lavrov@tspu.edu.ru}
\\[0.5cm]
\it ${}^a$Department of Theoretical Physics,\\
\it Tomsk State Pedagogical University,\\
\it Tomsk 634041, Russia\\[0.3cm]
\it ${}^b$Laboratory of Mathematical Physics,\\
\it Tomsk Polytechnic University,\\
\it Tomsk 634034, Russia\\[0.3cm]
\it ${}^c$Department of Mathematical Analysis,\\
\it Tomsk State Pedagogical University,\\
\it Tomsk 634041, Russia}
\date{}

\maketitle
\thispagestyle{empty}

\begin{abstract}
We develop the BRST approach to Lagrangian construction for the
massive integer higher spin fields in an arbitrary dimensional AdS
space. The theory is formulated in terms of auxiliary Fock space.
Closed nonlinear symmetry algebra of higher spin bosonic theory in
AdS space is found and method of deriving the BRST operator for such
an algebra is proposed. General procedure of Lagrangian construction
describing the dynamics of bosonic field with any spin is given on
the base of the BRST operator. No off-shell constraints on the
fields and the gauge parameters are used from the very beginning. As
an example of general procedure, we derive the Lagrangians for
massive bosonic fields with spin 0, 1 and 2 containing total set of
auxiliary fields and gauge symmetries.
\end{abstract}

\section{Introduction}

The various aspects of higher spin field theory attract much
attention in modern theoretical physics (see reviews
\cite{reviews}). At present, research in this area is devoted to
development of general methods allowing to construct Lagrangian
formulations, to study the specific properties, to clarify the
possible underlying structures of such a theory, to see the
relations with superstring theory and to find the ways leading
to description of interacting higher spin fields (see e.g.
\cite{We,Deser,Metsaev-m,Metsaev-2m,Zinoviev,Klishevich,
Heslop,Gates} for recent progress in massive and
\cite{Vasiliev,Metsaev-0,Sezgin,Sagnotti,Rajan,Bonelli,Bekaert,Sorokin,Barnich}
for recent progress in massless higher spin theories).

The present paper is devoted to formulating the general approach for
deriving the Lagrangians of massive higher spin fields in AdS space
of arbitrary dimension. The approach is based on development of BRST
construction in higher spin field theory which automatically allows
to obtain gauge invariant Lagrangian describing a consistent
dynamics of higher spin fields. To be more precise we use BRST-BFV
construction or BFV construction \cite{BFV} (see also the reviews
\cite{bf})\footnote{Following a tradition accepted in string field
theory and massless higher spin field theory we call BRST-BFV
construction as BRST construction.}.

BFV construction was originally developed for quantization of gauge
theories. The gauge theories are formulated in phase space and
characterized by first class constraints $T_{a}$ satisfying the
involution relations\footnote{We do not concern here the
unclosed gauge algebras.} in terms of Poisson brackets $\{T_{a},T_{b}\} =
f^{c}_{ab}T_{c}$ with structure functions $f^{c}_{ab}$. Basic notion
of BFV construction is nilpotent BFV charge $Q$ defined as follows
\begin{eqnarray}
\label{BFV} Q = \eta^{a}T_a +
\frac{1}{2}\eta^{b}\eta^{a}f^{c}_{ab}{\cal P}_{c} + \ldots\;,
\end{eqnarray}
where $\eta^{a}$ and ${\cal P}_{a}$ are canonically conjugate ghost
variables. The dots mean the extra terms which in principle should
be added in order to get the nilpotent charge for the case when the
structure functions depend on phase coordinates of the theory. After
quantization the BFV charge $Q$ becomes nilpotent Hermitian
operator\footnote{Following the accepted tradition we call this
operator as BRST operator.} in extended space of states $|\Psi\rangle$
containing the ghosts. Subspace of physical states is defined by the
equation $Q|\Psi\rangle  =0$ up to the transformation $|\Psi'\rangle
= |\Psi\rangle + Q|\Lambda\rangle$ which is gauge transformation in
the approach under consideration. It is proved that in subspace of
physical states an unitary $S$-matrix exists \cite{BFV}. We
emphasize that initial point of this construction is a classical
Lagrangian formulation of the gauge theory.

One of the problems of the higher spin field theory is its Lagrangian
formulation. As was pointed out in the pioneer paper \cite{PF} the
Lagrangian dynamics of higher spin fields demands to use, except a
basic filed with given spin $s$, the auxiliary fields with less spins.
Application of BRST approach to higher spin filed theory in
principle allows to describe its dynamics. The matter is that the
dynamics of gauge theory is completely concentrated into its
constraints. Since the BRST charge is constructed on the base of the
constraints, it includes whole information about dynamics. Therefore
if we are able to express the Lagrangian in terms of BRST charge, we
automatically obtain the consistent gauge invariant Lagrangian
formulation of the theory with all auxiliary fields. Namely such a
situation is realized in free higher spin field models
\cite{BRST-AdS,0410215,0505092,0511276,0603212}
(see also the early application of BRST approach to
interacting  higher spin filed theory in \cite{Beng}).

Use of BRST construction in higher spin filed theory differs from
its use for quantization in the following
points\footnote{Application of BRST construction in higher spin field
theory is analogous in some aspects to application of BRST
construction in string field theory \cite{W} (see also the reviews
\cite{SFT}). Relations of higher spin field theory to string field
theory are discussed in \cite{Ouvry}.}. Classical Lagrangian
formulation of the theory is unknown from the very beginning,
moreover the main problem of the theory is to find such a
formulation. Hence, classical system of constraints is also unknown
and classical BRST charge can not be written. The only we know are the
conditions on the fields defining irreducible representations of
Poincare or AdS group with given mass and spin. These conditions are
realized as some operators in auxiliary Fock space and interpreted
as first class operator constraints. Then ones demand that an
algebra of these operators in terms of their commutators should be
closed. This condition allows us to find a complete system of
operators forming the closed algebra. As a result, from the very
beginning we have the operators in Fock space and their
commutators, the formulation looks like quantum although it is still
a classical theory. Moreover, there are no any guaranties that the
commutator algebra is a Lie algebra. As we will see further, the
algebra in the case under consideration is nonlinear and analogous
to $W_3$ algebra (see e.g \cite{Sevrin}). Besides, it turns out that some
of the operators closing the algebra can not be interpreted as
constraints. Thus, application of BRST construction in higher spin
field theory faces many principal problems.

In this paper we develop the gauge invariant approach to deriving
the Lagrangian for massive higher spin fields in AdS space of
arbitrary dimension. This approach is based on BRST construction and
automatically yields a gauge invariant Lagrangian for field with
any spin containing a complete set of auxiliary fields. As it should
be, the gauge invariance of massive theory means that the
corresponding Lagrangian formulation includes the Stueckelberg
fields. The approach under consideration is a generalization of
works \cite{0410215,0505092,0511276,0603212} on Lagrangian
construction of higher spin fields in arbitrary dimensional
Minkowski space. The BRST approach to higher spin fields in AdS
space was first used in \cite{BRST-AdS}, where the bosonic massless
fields were considered. It was proved \cite{BRST-AdS} that in four
dimensions, obtained Lagrangian theory is reduced, after elimination
of all the auxiliary fields, to the Fronsdal Lagrangian
\cite{Fronsdal}. In this paper we develop a general method of
Lagrangian construction for massive higher spin bosonic fields in
AdS space and show that in massless limit it yields results
presented in \cite{BRST-AdS}

The paper is organized as follows. In Section 2 we introduce an
auxiliary Fock space, which is a base of BRST construction for
higher spin fields, and describe the differential operators
corresponding to constraints. These constraints define an
irreducible massive integer spin representation of the AdS group. In
Section 3 we construct a set of operators forming a closed algebra
in terms of commutators. It is shown that this algebra is nonlinear
(quadratic algebra analogous to $W_3$ algebra) and includes two
operators which are not constraints. In Section 4 we discuss an
extension of the algebra by means of generalization of the method
proposed in \cite{0410215,0505092,0511276,0603212}. This
generalization demands a deformation of the algebra and construction
of a representation for the deformed algebra. Such a representation
is found so that the new expressions for the operators consist of
two parts: the initial expression of the operator plus an additional
part. The additional parts for the operators are explicitly derived
in Section 5. Then in Section 6 we discuss the construction of BRST
operator for the deformed nonliner algebra. After this in
Section~\ref{Lagr-constr} we derive the Lagrangian describing
propagation of massive bosonic field of arbitrary fixed spin in AdS
space. There it is also shown that this theory is a gauge one and
the gauge transformations are written down. In
Section~\ref{examples} we illustrate the procedure of Lagrangian
construction by finding the gauge invariant Lagrangians for massive
spin-0, spin-1, and spin-2 fields and their gauge transformations in
explicit form. Section~\ref{summary} summarizes the obtained
results. Finally, in Appendix~\ref{Fock-Ap} we give general
differential calculus in auxiliary Fock space on a gravitational
background. Appendix~\ref{Ap-AddParts} is devoted to obtaining of a
representation of the operator algebra given in Table~\ref{table'}
in terms of creation and annihilation operators and in
Appendix~\ref{proof} we prove that the constructed Lagrangian
reproduces the correct conditions on the field defining the
irreducible representation of the AdS group.

\section{Auxiliary Fock space for higher spin fields in AdS
space-time. }\label{Fock}

Massive integer spin-$s$ representation of the AdS group is realized
in space of totally symmetric tensor fields
$\Phi_{\mu_1\ldots\mu_s}(x)$ satisfying the following conditions
(see e.g. \cite{Metsaev-m})
\begin{eqnarray}
\label{Eq-0} && \bigl(\nabla^2+r[s^2+s(d-6)+6-2d]+m^2\bigr)
\Phi_{\mu_1\mu_2\ldots\mu_s}(x) =0,
\\
&& \nabla^{\mu_1} \Phi_{\mu_1\mu_2\ldots\mu_s}(x) =0,
\\
&& g^{\mu_1\mu_2} \Phi_{\mu_1\mu_2\ldots\mu_s}(x) =0,
\label{Eq-2}
\end{eqnarray}
where $r=\frac{R}{d(d-1)}$ with $R$ being the scalar curvature and
$d$ being the dimension of the space-time. Our purpose is to
construct a Lagrangian which reproduces these constraints as the
consequences of the equations of motion.

In order to avoid an explicit manipulation with a number of indices
it is convenient to introduce auxiliary Fock space generated by
creation and annihilation operators with tangent space indices
($a,b=0,1,\ldots,d-1$)
\begin{eqnarray}
[a_a,a^+_b]=-\eta_{ab},
\qquad
\eta_{ab}=diag(+,-,\ldots,-).
\end{eqnarray}
An arbitrary vector in this Fock space has the form
\begin{eqnarray}
\label{PhysState}
|\Phi\rangle
&=&
\sum_{s=0}^{\infty}\Phi_{a_1\ldots\,a_s}(x)\,
a^{+a_1}\ldots\,a^{+a_s}|0\rangle
=
\sum_{s=0}^{\infty}\Phi_{\mu_1\ldots\,\mu_s}(x)\,
a^{+\mu_1}\ldots\,a^{+\mu_s}|0\rangle
,
\end{eqnarray}
where
$a^{+\mu}(x)=e^\mu_a(x)a^{+a}$, $a^\mu(x)=e^\mu_a(x)a^a$,
with $e^\mu_a(x)$ being the vielbein.
It is evident that
\begin{eqnarray}
[a_\mu, a_\nu^+]=-g_{\mu\nu}.
\end{eqnarray}
We call the vector (\ref{PhysState}) as basic vector. The fields
$\Phi_{\mu_1\ldots\mu_s}(x)$ are the coefficient functions of the
vector $|\Phi\rangle$ and its symmetry properties are stipulated by
the properties of the product of the creation operators. We also
suppose the standard relation
\begin{eqnarray}
\nabla_\mu e^a_\nu
=0.
\end{eqnarray}

We want to realize the constraints (\ref{Eq-0})--(\ref{Eq-2}) as
some constraints on the vectors $|\Phi\rangle$. To do that ones
introduce a special derivative operator acting on the vectors
$|\Phi\rangle$\footnote{Another (equivalent) form of the derivative
operators acting on vectors of Fock space and differential calculus
based on such operators are discussed in Appendix~\ref{Fock-Ap}.}
\begin{eqnarray}
D_\mu=\partial_\mu-\omega_\mu^{ab}a_a^+a_b,
&\qquad&
D_\mu|0\rangle=0.
\label{D_mu}
\end{eqnarray}
Then one can show that
\begin{eqnarray}
[D_\mu,
\Phi_{\mu_1\ldots\mu_s}]=(\partial_\mu\Phi_{\mu_1\ldots\mu_s}),
\end{eqnarray}
\begin{align}
&
\Bigl[D_\mu,a^{a+}\Bigr]=\omega_\mu^{ba}a_b^+,
&&
\Bigl[D_\mu,a^a\Bigr]=-\omega_\mu^{ab}a_b,
\\
&
\Bigl[D_\mu,a^{+\nu}\Bigr]=-\Gamma^\nu_{\mu\lambda}a^{+\lambda},
&&
\Bigl[D_\mu,a^{\nu}\Bigr]=-\Gamma^\nu_{\mu\lambda}a^{\lambda},
\\
&
\Bigl[D_\mu,a^+_\nu\Bigr]=\Gamma^\lambda_{\mu\nu}a^+_\lambda,
&&
\Bigl[D_\mu,a_\nu\Bigr]=\Gamma^\lambda_{\mu\nu}a_\lambda,
\end{align}
where we suppose that $\partial_\mu{}a_a=\partial_\mu{}a_a^+=0$.
The operator $D_\mu$ acts on the vectors (\ref{PhysState}) as
\begin{eqnarray}
D_\mu|\Phi\rangle
&=&
\sum_{s=0}^{\infty}(\nabla_\mu\Phi_{a_1\ldots\,a_s}(x))\,
a^{+a_1}\ldots\,a^{+a_s}|0\rangle
=
\sum_{s=0}^{\infty}(\nabla_\mu\Phi_{\mu_1\ldots\,\mu_s}(x))\,
a^{+\mu_1}\ldots\,a^{+\mu_s}|0\rangle
,
\end{eqnarray}
where $\nabla_\mu$ is the covariant derivative acting on tensor
fields with tangent and space-time indices by standard rule.

For latter
purposes it is useful to state the relations
\begin{eqnarray}
g^{\mu\nu}(D_\mu D_\nu-\Gamma^\sigma_{\mu\nu}D_\sigma)|\Phi\rangle
&=&
\sum_{n=0}^\infty (\nabla^\mu\nabla_\mu \Phi_{\mu_1\cdots\mu_n}(x))
  a^{\mu_1+}\cdots a^{\mu_n+}|0\rangle,
\\
\bigl[ D_\nu, D_\mu \bigr]
&=&
{\cal R}_{\mu\nu}{}^{ab}a_a^+a_b
=R_{\mu\nu\rho\tau}a^{\rho+}a^{\tau},
\end{eqnarray}
with ${\cal R}_{\mu\nu}{}^{ab}$, $R_{\mu\nu\rho\tau}$ being the
curvatures (\ref{F23}), (\ref{F20}). Here and further we use
following convention for AdS curvature tensor
$R_{\mu\nu\alpha\beta}=g_{\mu\lambda}R^{\lambda}_{\;\;\nu\alpha\beta}=
r(g_{\mu\alpha} g_{\nu\beta}-g_{\mu\beta}g_{\nu\alpha})$.

The operator $D_{\mu}$ is used for realization of the constraints in
space of vectors $|\Phi\rangle$. Let us define the following
operators
\begin{eqnarray}
\label{l0-kaz} l_0&=&
D^2+m^2+r[X-4g_0+6-{\textstyle\frac{d(d-4)}{4}}\;] ,
\\
&&
D^2 = g^{\mu\nu}(D_\mu D_{\nu}-\Gamma^\sigma_{\mu\nu}D_\sigma),
\\
&&
X=g_0^2-2g_0-4l_2^+l_2,
\end{eqnarray}
\begin{align}
& \label{l1}
l_1=-ia^\mu D_\mu,
&&
l_1^+=-ia^{\mu+} D_\mu,
\\
&
\label{l2}
l_2={\textstyle\frac{1}{2}}\,a^\mu a_\mu,
&&
l_2^+={\textstyle\frac{1}{2}}\,a^{\mu+} a_\mu^+,
\\
& g_0=-a_\mu^+ a^\mu+{\textstyle\frac{d}{2}} .
\label{g0}
\end{align}
Using the above relations, one can see that if the following
constraints
\begin{eqnarray}
l_0|\Phi\rangle=
l_1|\Phi\rangle=
l_2|\Phi\rangle=
0
\label{constr}
\end{eqnarray}
are fulfilled then each component $\Phi_{\mu_1\ldots\,\mu_s}(x)$ of
(\ref{PhysState}) obeys the conditions describing spin-$s$ field in
AdS space (\ref{Eq-0})--(\ref{Eq-2}). Therefore if the conditions
(\ref{constr}) are fulfilled then vector (\ref{PhysState}) will
describe the fields with arbitrary integer spin $s$ in AdS space.

Let us turn to the algebra generated by operators $l_0$, $l_1$,
$l_2$.

\section{Algebra generated by the constraints}

We develop an approach to Lagrangian formulation for massive higher
spin fields based on the following idea. We treat the conditions
(\ref{Eq-0})--(\ref{Eq-2}) or the equivalent conditions
(\ref{constr}) as the constraints in some unknown yet Lagrangian
gauge theory. Our purpose is to restore a Lagrangian leading to
given set of first class constraints. The most efficient way to
realize this idea is to use general BRST method.

Basic notion of BRST method is Hermitian BRST operator which is
constructed on the base of a set of first class constraints. To get
Hermitian BRST operator, the set of constraints should be invariant
under Hermitian conjugation. In the case under consideration, the
operator $l_0$ is Hermitian, but the operators $l_{1}, l_{2}$ are
not Hermitian. Therefore, to get a set of operators invariant under
hermitian conjugation we should add two more operators $l_1^+$
(\ref{l1}), $l_2^+$ (\ref{l2}) to the operators $l_0$, $l_1$, $l_2$.
As a result, the set of operators $l_0$, $l_1$, $l_1^+$, $l_2$,
$l_2^+$ is invariant under Hermitian conjugation. However, it is
clear that the operators $l_1^+$, $l_2^+$ can not be interpreted as
constraints on the vectors (\ref{PhysState}) on equal footing with
the operators $l_{0}$, $l_{1}$, $l_{2}$. Indeed, taking Hermitian
conjugation of (\ref{constr}) we see that they (together with $l_0$)
are constraints on the bra vector
\begin{eqnarray}
\langle\Phi|l_0
=
\langle\Phi|l_1^+
=
\langle\Phi|l_2^+
=0.
\label{bra-constr}
\end{eqnarray}
Nevertheless we will use the operators $l^{+}_{1}$, $l^{+}_{2}$ for
BRST construction and show that they do not contribute to equations
of motion for basic vector (\ref{PhysState}) obtained from final
Lagrangian. This occurs due to these operators are multiplied on the
annihilation ghost operators in BRST operator (see relations
(\ref{Q-ini}) and (\ref{ghostvac}), some details are discussed in
\cite{0505092}).

Algebra of the operators $l_0$, $l_1$, $l_1^+$, $l_2$, $l_2^+$ is
open in terms of commutators of these operators. To close the
algebra we add to it all the operators generated by the commutators
of $l_0$, $l_1$, $l_1^+$, $l_2$, $l_2^+$.

Before proceeding in this way let us introduce a more general
operator $\tilde{l}_0$ instead of operator $l_0$
\begin{eqnarray}
\tilde{l}_0
&=&
D^2+m^2+(\alpha-1)r{\textstyle\frac{d(d-4)}{4}}
+(\beta-1)rX+\gamma rg_0
\label{l0}
\end{eqnarray}
and use this operator $\tilde{l}_0$ as a constraint. The
operator $\tilde{l}_0$ depends on real parameters ${\alpha}$,
${\beta}$, ${\gamma}$. It is evident that $l_0$ (\ref{l0-kaz}) is
obtained from $\tilde{l}_0$ at $\alpha=\frac{24}{d(d-4)}$,
$\beta=2$, $\gamma=-4$. We shall show that the correct on-shell
condition (\ref{Eq-0}) can be reproduced at any values of $\alpha$,
$\beta$, $\gamma$, besides the expressions of the Lagrangian
(\ref{Lagr}) will be simpler for $\tilde{l}_0$ at some specific
values ${\beta}$, ${\gamma}$, which do not coincide with $l_0$ (see
Appendix~\ref{proof}). Also we note that the case of massless
bosonic higher spin fields in AdS space considered in
\cite{BRST-AdS} corresponds to the following choice of the
coefficients $m=\alpha=\gamma=0$, $\beta=2$.

Now let us close the algebra generated by the operators
$\tilde{l}_0$, $l_1$, $l_1^+$, $l_2$, $l_2^+$. As a result we obtain
two more Hermitian operators: $g_0$ (\ref{g0}) and
\begin{eqnarray}
l&=&m^2+\alpha \frac{d(d-4)}{4}\, r.
\label{M-op}
\end{eqnarray}

The algebra of operators (\ref{l1})--(\ref{g0}), (\ref{l0}),
(\ref{M-op}) is closed and it is given in Table~\ref{table}, where
\begin{eqnarray}
[l_1,\tilde{l}_0]&=&
\beta r(4l_1^+l_2+2g_0l_1-l_1)
+\gamma r l_1
\label{l1l0}
,
\\
{}
[\tilde{l}_0,l_1^+]&=&
\beta r(4l_2^+l_1+2l_1^+g_0-l_1^+)
+\gamma r l_1^+
\label{l0l1+}
,
\\
{}
[l_1,l_1^+]&=&
\tilde{l}_0-l
+(2-\beta)rX-\gamma rg_0
.
\label{l1l1+}
\end{eqnarray}
\begin{table}
\begin{eqnarray*}
&
\begin{array}{||l||r|r|r|r|r||r|c||}\hline\hline\vphantom{\biggm|}
       &\tilde{l}_0&l_1&l_1^+&l_2&l_2^+&g_0&\quad l\quad\\
\hline\hline\vphantom{\biggm|}
\tilde{l}_0
          &0&-(\ref{l1l0})&(\ref{l0l1+})
          &-2\gamma rl_2&2\gamma rl_2^+&0&0
              \\
\hline\vphantom{\biggm|}
l_1
          &(\ref{l1l0})&0&(\ref{l1l1+})&0&-l_1^+&l_1&0 \\
\hline\vphantom{\biggm|}
l_1^+
          &-(\ref{l0l1+})&-(\ref{l1l1+})&0&l_1&0&-l_1^+&0 \\
\hline\vphantom{\biggm|}
l_{2}
          &2\gamma rl_2&0&-l_1&0&g_0&2l_2&0\\
\hline\vphantom{\biggm|}
l_{2}^+
          &-2\gamma rl_2^+&l_1^+&0&-g_0&0&-2l_2^+&0\\
\hline\hline\vphantom{\biggm|}
g_{0}
          &0&-l_1&l_1^+&-2l_2&2l_2^+&0&0\\
\hline\vphantom{\biggm|}
l
          &0&0&0&0&0&0&0\\
\hline\hline
\end{array}
\end{eqnarray*}
\caption{Algebra of the operators
(\ref{l1})--(\ref{g0}), (\ref{l0}), (\ref{M-op})}
\label{table}
\end{table}
In this Table and in the subsequent ones
the first arguments of the commutators are listed in
the first column, the second arguments are listed in the upper row.
The algebra corresponding to Table 1 is a base for massive integer
higher spin filed Lagrangian construction in AdS space. We will call
this algebra as massive integer higher spin symmetry algebra in AdS
space. We want to emphasize here two points. First, the operator $l$
commutes with all other operators and therefore it plays a role of
central charge. Second, due to (\ref{l1l0})--(\ref{l1l1+}), the
given algebra is non-linear, the right hand sides of these
commutators are quadratic forms in operators.

Since operators $g_0$ and $l$ are obtained as commutators of a
constraint on the ket vector (\ref{constr}) with a constraint on the
bra vector (\ref{bra-constr}), these operators $g_0$ and $l$ can not
be considered as constraints neither in the space of bra-vectors nor
in the space of ket-vectors.

One can show that a straightforward use of BRST construction as if
all the operators are the first class constraints doesn't lead to
the proper equations (\ref{constr}) for any spin (see e.g.
\cite{0410215,0505092,0511276,0603212} for the cases of higher spin
fields in flat space). This happens because among the above
hermitian operators there are operators which are not
constraints ($g_0$ and $l$ in the case under consideration) and
they bring two more equations (in addition to (\ref{constr})) on
the physical field (\ref{PhysState}).
Thus we must somehow get rid of these supplementary equations.

Method of avoiding the supplementary equations consists in
constructing new enlarged expressions for the operators of the algebra
given in Table~\ref{table} so that the hermitian operators which are
not constraints will be zero.

\section{Constructing the deformations of the
symmetry algebra}\label{method}

Our purpose here is to construct the new enlarged operators
satisfying some deformed algebra so that the new operators $g_0$ and
$l$ become zero or (as was shown in \cite{0505092}) they contain
arbitrary parameters which then defined from the condition that the
supplementary equations do not give more restrictions on the basic
vectors (\ref{PhysState}) in addition to (\ref{constr}).

The method we apply in this paper for constructing the enlarged
expressions for the operators is a generalization of the method used
in our papers \cite{0410215,0505092,0511276,0603212} where the
operator algebras are Lie algebras.

In case of Lie algebras we acted as follows. We enlarged the
representation space introducing new creation and annihilation
operators and constructed a new representation of the corresponding
operators so that the expression for any operator was a sum of two
parts: the initial expression for the operator plus a specific part
which depends on the new creation and annihilation operators only.
As a result in this new representation the operators which are not
constraints were zero or contained arbitrary parameters whose values
would define later\footnote{Introducing the new creation and
annihilation operators in BRST construction for higher spin field
theory is analogous to a conversion procedure (see e.g.
\cite{conversion}) which is used for quantization of constraint
systems with second class constraints.}. One more requirement was
that the vector in the enlarged space (including the ghost fields)
should be independent of the ghost fields corresponding to the
operators which are not constraints
\cite{0410215,0505092,0511276,0603212}.

The generalization of the above method to the case of non-linear symmetry
algebra given by Table~\ref{table} is based on deformation of
algebra of the enlarged operators in comparison with the corresponding
initial algebra.

Ones describe the method of deformation. Let us denote all the
operators of the algebra given in Table~\ref{table} as $l_i$. Then
the structure of the algebra looks like\footnote{The method can
easily be generalized to the case when the algebra has the structure
\begin{displaymath}
[\,l_i,l_j]=
f_{ij}^kl_k+f_{ij}^{km}l_kl_m
+\cdots+
f_{ij}^{k_1\ldots\,k_N}l_{k_1}\ldots\,l_{k_N}
\end{displaymath}
where all $f$'s are constants. }
\begin{eqnarray}
[\,l_i,l_j]&=&
f_{ij}^kl_k+f_{ij}^{km}l_kl_m,
\label{alg-ini}
\end{eqnarray}
where $f_{ij}^k$, $f_{ij}^{km}$ are constants. In case of
$f_{ij}^{km}=0$ we have a Lie algebra and the described method will
be reduced to the method used in the flat space case
\cite{0410215,0505092,0511276,0603212}. As in the case of a Lie
algebra we enlarge the representation space by introducing the
additional creation and annihilation operators and construct the new
operators of the algebra $l_i\to{}L_i$
\begin{eqnarray}
L_i&=&l_i+l_i',
\label{Li}
\end{eqnarray}
where $l_i'$ are the part of the operator which depends on the new
creation and annihilation operators only (and constants of the
theory like the mass $m$ and the curvature). It is evident that if
we fix the structure of the new operators (\ref{Li}) then we are not
able to preserve the algebra (\ref{alg-ini}). Therefore there are
two possibilities. The first one is to reject (\ref{Li}) and the
second one is to deform the algebra (\ref{alg-ini}). We choose the
second possibility and demand that the new operators $L_i$
(\ref{Li}) are in involution relations
\begin{eqnarray}
[\,L_i,L_j]&\sim& L_k.
\label{invL}
\end{eqnarray}
Since $[\,l_i,l_j']=0$ we have
\begin{eqnarray}
[\,L_i,L_j]&=&
[\,l_i,l_j]+[\,l_i',l_j']=
\nonumber
\\
&=&
f_{ij}^kL_k
-(f_{ij}^{km}+f_{ij}^{mk})l_m'L_k+f_{ij}^{km}L_kL_m
-f_{ij}^kl_k'+f_{ij}^{km}l_m'l_k'
+[\,l_i',l_j']
.
\end{eqnarray}
Then in order to provide (\ref{invL}) the last three terms must
be canceled. Therefore we put
\begin{eqnarray}
[\,l_i',l_j']
&=&
f_{ij}^kl_k'-f_{ij}^{km}l_m'l_k'
\label{addal}
\end{eqnarray}
and as a result we have the deformed algebra for the enlarged operators
\begin{eqnarray}
[\,L_i,L_j]&=&
f_{ij}^kL_k
-(f_{ij}^{km}+f_{ij}^{mk})l_m'L_k+f_{ij}^{km}L_kL_m
.
\label{alg-enl}
\end{eqnarray}
Thus we see that the algebra (\ref{alg-enl}) of the enlarged
operators $L_i$ is deformed in comparison with the algebra
(\ref{alg-ini}) of the initial operators $l_i$. The second term in
this algebra shows that some of the structure functions are not the
constants and depend on the new creation and annihilation operators
by means of $l_{m}'$. In case of Lie algebra $f_{ij}^{km}=0$ the
algebras of the initial operators $l_i$, of the additional parts
$l_i'$, and of the enlarged operators $L_i$ coincide. Namely this
situation takes place in the flat space
\cite{0410215,0505092,0511276,0603212}.

Before we go further, ones discuss some details concerning the
construction of the algebra (\ref{alg-enl}). First, we should find
the operators (additional parts) $l_i'$ satisfying the algebra
(\ref{addal}). Second, these additional parts corresponding to the
operators which are not constraints ($g_0'$ and $l'$ in the case
under consideration) should contain an arbitrary parameter or vanish
in sum with their initial expression ($g_0$ and $l$ in the case
under consideration). Third, the additional part corresponding to
the Hermitian constraint $\tilde{l}_0'$ should contain an arbitrary
parameter which value will be defined from the condition of
reproducing the correct conditions (\ref{constr}). After the
additional parts $l_i'$ will be calculated we proceed as follows:
ones construct BRST operator as if all the operators $L_i$ are the
first class constraints and the vector in the enlarged space
(including the ghost fields) must
be independent of the ghost fields corresponding to the operators
which are not constraints.

In next Section we study the algebra (\ref{addal}) of the additional
parts of the operators $l_i'$ and find their explicit expressions.

\section{Constructing the additional parts of the operators}\label{addparts}

The procedure described in Section 4  in the case under
consideration leads to algebra (\ref{addal}) for the operators of
the additional parts in the form given by
 Table~\ref{table'}
\begin{table}
\begin{eqnarray*}
&
\begin{array}{||l||r|r|r|r|r||r|c||}\hline\hline\vphantom{\biggm|}
       &\tilde{l}_0'&l_1'&l_1^{\prime+}&l_2'&l_2^{\prime+}&g_0'&
       \quad l'\quad\\
\hline\hline\vphantom{\biggm|}
\tilde{l}_0'
          &0&-(\ref{l1l0'})&(\ref{l0l1+'})
          &-2\gamma rl_2'&2\gamma rl_2^{\prime+}&0&0
              \\
\hline\vphantom{\biggm|}
l_1'
          &(\ref{l1l0'})&0&(\ref{l1l1+'})&0&-l_1^{\prime+}&l_1'&0
          \\
\hline\vphantom{\biggm|}
l_1^{\prime+}
          &-(\ref{l0l1+'})&-(\ref{l1l1+'})
          &0&l_1'&0&-l_1^{\prime+}&0 \\
\hline\vphantom{\biggm|}
l_2'
          &2\gamma rl_2'&0&-l_1'&0&g_0'&2l_2'&0\\
\hline\vphantom{\biggm|}
l_2^{\prime+}
          &-2\gamma rl_2^{\prime+}&l_1^{\prime+}&0
          &-g_0'&0&-2l_2^{\prime+}&0\\
\hline\hline\vphantom{\biggm|}
g_0'
          &0&-l_1'&l_1^{\prime+}&-2l_2'&2l_2^{\prime+}&0&0\\
\hline\vphantom{\biggm|}
l'
          &0&0&0&0&0&0&0\\
\hline\hline
\end{array}
\end{eqnarray*}
\caption{Algebra of the additional parts for the operators}
\label{table'}
\end{table}
where
\begin{eqnarray}
[l_1',\tilde{l}_0']
&=&
-\beta r(4l_1^{\prime+}l_2'+2g_0'l_1'-l_1')
+\gamma rl_1'
,
\label{l1l0'}
\\
{}
[\tilde{l}_0',l_1^{\prime+}]
&=&
-\beta r (4l_2^{\prime+}l_1'+2l_1^{\prime+}g_0'-l_1^{\prime+})
+\gamma rl_1^{\prime+}
,
\label{l0l1+'}
\\{}
[l_1',l_1^{\prime+}]
&=&
\tilde{l}_0'-l'-\gamma rg_0'
+(\beta-2)r(g_0^{\prime2}-2g_0'-4l_2^{\prime+}l_2')
.
\label{l1l1+'}
\end{eqnarray}

Ones point out that in the case of massless fields considered in
\cite{BRST-AdS} $m=\alpha=\gamma=0$, $\beta=2$, and therefore there
was possible to construct the additional parts so that
$\tilde{l}_0'=l'=l_1'=l_1^{\prime+}=0$.

In the more general case considered here the explicit expressions
for the additional parts can be calculated with the help of the
method described in\footnote{To be completed a detailed calculation
of the additional parts is given in Appendix~\ref{Ap-AddParts}.}
\cite{0206027,0410215}. As a result we get
\begin{eqnarray}
\label{l0'}
\tilde{l}_0' &=& m_0^2+\gamma r(b_1^+b_1+2b_2^+b_2) - \beta r
(2h-1)\,b_1^+ \sum_{k=0}^{\infty} \left[\frac{-8r}{m_1^2}\right]^k
\frac{b_2^{+k}\,b_1^{2k+1}}{(2k+1)!} \nonumber
\\
&& {} - 2\beta r \, b_1^{+2} \sum_{k=0}^{\infty}
\left[\frac{-8r}{m_1^2}\right]^k
\frac{b_2^{+k}\,b_1^{2k+2}}{(2k+2)!} + \frac{1}{2}\beta M^2
\sum_{k=1}^{\infty} \left[\frac{-8r}{m_1^2}\right]^k
\frac{b_2^{+k}\,b_1^{2k}}{(2k)!}
,
\end{eqnarray}
\begin{eqnarray}
l_1'
&=&
-m_1b_1^+b_2
+
m_1b_1^+
\frac{2h-1}{4}
\sum_{k=1}^{\infty}
\left[\frac{-8r}{m_1^2}\right]^k
\frac{(b_2^+)^{k-1}\,b_1^{2k}}{(2k)!}
\nonumber
\\
&& {} + \frac{1}{2} \, m_1b_1^{+2} \sum_{k=1}^{\infty}
\left[\frac{-8r}{m_1^2}\right]^k
\frac{(b_2^+)^{k-1}\,b_1^{2k+1}}{(2k+1)!} + \frac{M^2}{m_1}
\sum_{k=0}^{\infty} \left[\frac{-8r}{m_1^2}\right]^k
\frac{b_2^{+k}\,b_1^{2k+1}}{(2k+1)!}
,
\end{eqnarray}
\begin{eqnarray}
l_2'
&=&
(b_2^+b_2+b_1^+b_1+h)b_2
-
\frac{2h-1}{4}
\,
b_1^+
\sum_{k=1}^{\infty}
\left[\frac{-8r}{m_1^2}\right]^k
\frac{(b_2^+)^{k-1}\,b_1^{2k+1}}{(2k+1)!}
\nonumber
\\
&& {} -\frac{1}{2} \, b_1^{+2} \sum_{k=1}^{\infty}
\left[\frac{-8r}{m_1^2}\right]^k
\frac{(b_2^+)^{k-1}\,b_1^{2k+2}}{(2k+2)!} - \frac{M^2}{m_1^2}
\sum_{k=0}^{\infty} \left[\frac{-8r}{m_1^2}\right]^k
\frac{b_2^{+k}\,b_1^{2k+2}}{(2k+2)!} ,
\end{eqnarray}
\begin{eqnarray}
l_1^{\prime+}
&=&
m_1b_1^+,
\\
l_2^{\prime+}
&=&
b_2^+,
\\
g_0'&=&
b_1^+b_1+2b_2^+b_2+h,
\\
l'&=&-m^2-\alpha{\textstyle\frac{d(d-4)}{4}}\,r,
\label{M'}
\end{eqnarray}
where we have denoted
\begin{eqnarray}
M^2&=&
m^2+m_0^2+\alpha{\textstyle\frac{d(d-4)}{4}}\,r
+(\beta-2)h(h-2)r-\gamma hr
.
\label{Mass}
\end{eqnarray}
In the above expressions  $h$ is a dimensionless arbitrary constant,
$m_0$ and $m_1$ are the arbitrary constants with dimension of mass.
For constructing the additional parts (\ref{l0'})--(\ref{M'}) we
have introduced, in accordance with method given in Section 4, two
pairs of new bosonic creation and annihilation operators satisfying
the standard commutation relations
\begin{eqnarray}
[b_1,b_1^+]=[b_2,b_2^+]=1.
\end{eqnarray}

The found additional parts of operators possess all the necessary
properties described in Section 4. The operators which are not
constraints give zero in sum with their initial expressions for the
operator ($l+l'=0$) or contain an arbitrary parameter ($g_0'$
contains an arbitrary parameter $h$) which value will be defined
later from the condition that they do not generate the extra
restrictions on the physical states. The additional part for the
Hermitian operator $\tilde{l}_0'$ also contains an arbitrary
parameter $m_0^2$ which value will be defined from the condition of
reproducing the correct conditions (\ref{constr}). The massive
parameter $m_1$ remains arbitrary, in particular it can be expressed
through the other massive parameters of the theory
\begin{eqnarray}
m_1=f(m,r)\neq0.
\label{m1}
\end{eqnarray}
For example one can put $m_1=m$ or $m_1=\lambda$, where $\lambda$ is
the inverse radius of the AdS space $\lambda^2=r$. This
arbitrariness does not affect on the equations for the basic vector
(\ref{PhysState}).

We note that the operators of the additional parts do not satisfy
the usual properties
\begin{eqnarray}
(\tilde{l}_0')^+\neq \tilde{l}_0',
\qquad
(l_1')^+\neq l_1^{\prime+},
\qquad
(l_2')^+\neq l_2^{\prime+}
\end{eqnarray}
if we use the standard rules of Hermitian conjugation for the new
creation and annihilation operators
\begin{eqnarray}
(b_1)^+=b_1^+,
&\qquad&
(b_2)^+=b_2^+.
\end{eqnarray}

To restore the proper Hermitian conjugation properties for the
additional parts we change the scalar product in the Fock space
generated by the new creation and annihilation operators
\begin{eqnarray}
\langle\Psi_1|\Psi_2\rangle_{new}
&=&
\langle\Psi_1|K|\Psi_2\rangle
\label{sprod}
\end{eqnarray}
with some unknown yet operator $K$ . This operators is defined from
the condition that all the operators must
have the proper Hermitian properties
with respect to
the new scalar product
\begin{align}
\label{H0}
&
\langle\Psi_1|K\tilde{l}_0'|\Psi_2\rangle=
\langle\Psi_1|(\tilde{l}_0)^+K|\Psi_2\rangle,
&&
\langle\Psi_1|Kg_0'|\Psi_2\rangle=
\langle\Psi_1|(g_0)^+K|\Psi_2\rangle,
\\
\label{H1}
&
\langle\Psi_1|Kl_1'|\Psi_2\rangle=
\langle\Psi_1|(l_1^{\prime+})^+K|\Psi_2\rangle,
&&
\langle\Psi_1|Kl_2'|\Psi_2\rangle=
\langle\Psi_1|(l_2^{\prime+})^+K|\Psi_2\rangle,
\\
&
\langle\Psi_1|Kl_1^{\prime+}|\Psi_2\rangle=
\langle\Psi_1|(l_1^{\prime})^+K|\Psi_2\rangle,
&&
\langle\Psi_1|Kl_2^{\prime+}|\Psi_2\rangle=
\langle\Psi_1|(l_2^{\prime})^+K|\Psi_2\rangle.
\label{H2}
\end{align}
The explicit expression for the operator $K$ can be found using the
method given in \cite{0101201,0410215, 0505092}. The calculations of
the operator $K$ are described in Appendix~\ref{Ap-AddParts}.

\section{The deformed algebra and BRST operator}\label{BRST}

In this section we find BRST operator and discuss the aspects
stipulated by nonlinearity of the operator algebra.

Construction of BRST operator is based on following general
principles (see the details in reviews \cite{bf}):

{\bf 1.} BRST operator $Q'$ is Hermitian, $Q^{\prime+}=Q'$, and
nilpotent, $Q^{\prime2}=0$.

{\bf 2.} BRST operator $Q'$ is built using a set of first class
constraints. In the case under consideration the operators
$\tilde{L}_{0}$,
$L_{1}$, $L_{1}^{+}$, $L_{2}$, $L_{2}^{+}$, $G_{0}$ are used as a
set of such constraints.

{\bf 3.} BRST operator $Q'$ satisfies the special initial condition
\begin{eqnarray}
Q'\Big|_{{\cal{}P}=0}
&=&
\eta_0\tilde{L}_0+\eta_1^+L_1+\eta_1L_1^+
+\eta_2^+L_2+\eta_2L_2^++\eta_{G}G_0
.
\label{Q-ini}
\end{eqnarray}
Here $\eta_0$, $\eta_1^+$, $\eta_1$, $\eta_2^+$, $\eta_2$, $\eta_G$
are ghost 'coordinates' with ghost number $gh({\eta})=+1$ and
${\cal{}P}_0$, ${\cal{}P}_1$, ${\cal{}P}_1^+$, ${\cal{}P}_2$,
${\cal{}P}_2^+$, ${\cal{}P}_G$ are their canonically conjugate ghost
'momenta' with ghost number $gh({\cal{}P})=-1$ satisfying the
anticommutation relations
\begin{eqnarray}
&&
\{\eta_0,{\cal{}P}_0\}=
\{\eta_G,{\cal{}P}_G\}=
\{\eta_1,{\cal{}P}_1^+\}=
\{\eta_1^+,{\cal{}P}_1\}=
\{\eta_2,{\cal{}P}_2^+\}=
\{\eta_2^+,{\cal{}P}_2\}=1.
\end{eqnarray}
Our purpose here is to construct such an operator $Q'$ in an explicit
form.

Let us turn to the algebra of the enlarged operators $L_i$.
Straightforward calculation of commutators (in accordance with
Section~\ref{method}) allows to to find this algebra in the
form\footnote{In what follows we forget about operator $l$, since
its enlarged expression is zero $l+l'=0$. See formulas (\ref{M-op})
and (\ref{M'}).} given in Table~\ref{Table},
\begin{table}
\begin{eqnarray*}
&
\begin{array}{||l||r|r|r|r|r||r||}\hline\hline\vphantom{\biggm|}
       &\tilde{L}_0&L_1&L_1^+&L_2&L_2^+&G_0\\
\hline\hline\vphantom{\biggm|}
\tilde{L}_0
          &0&-(\ref{L1L0})&(\ref{L0L1+})
          &-2\gamma rL_2&2\gamma rL_2^+&0
              \\
\hline\vphantom{\biggm|}
L_1
          &(\ref{L1L0})&0&(\ref{L1L1+})&0&-L_1^+&L_1 \\
\hline\vphantom{\biggm|}
L_1^+
          &-(\ref{L0L1+})&-(\ref{L1L1+})&0&L_1&0&-L_1^+ \\
\hline\vphantom{\biggm|}
L_{2}
          &2\gamma rL_2&0&-L_1&0&G_0&2L_2\\
\hline\vphantom{\biggm|}
L_{2}^+
          &-2\gamma rL_2^+&L_1^+&0&-G_0&0&-2L_2^+\\
\hline\hline\vphantom{\biggm|}
G_{0}
          &0&-L_1&L_1^+&-2L_2&2L_2^+&0\\
\hline\hline
\end{array}
\end{eqnarray*}
\caption{Algebra of the enlarged operators}
\label{Table}
\end{table}
where
\begin{eqnarray}
[L_1,\tilde{L}_0]&=&
(\gamma-\beta) r L_1
+4\beta rL_1^+L_2
-4\beta rl_1^{\prime+}L_2
-4\beta rl_2'L_1^+
\nonumber
\\
&&{}
+2\beta r G_0L_1
-2\beta rl_1'G_0
-2\beta rg_0'L_1
,
\label{L1L0}
\\
{}
[\tilde{L}_0,L_1^+]
&=&
(\gamma-\beta) r L_1^+
+4\beta rL_2^+L_1
-4\beta rl_2^{\prime+}L_1
-4\beta rl_1'L_2^+
\nonumber
\\
&&{}
+2\beta r L_1^+G_0
-2\beta r l_1^{\prime+}G_0
-2\beta r g_0'L_1^+
,
\label{L0L1+}
\\
{}
[L_1,L_1^+]
&=&
\tilde{L}_0-\gamma rG_0
+4(2-\beta)r(l_2^{\prime+}L_2+l_2'L_2^+)
\nonumber
\\
&&{}
-2(2-\beta)rg_0'G_0
+(2-\beta)r(G_0^2-2G_0-4L_2^+L_2)
.
\label{L1L1+}
\end{eqnarray}

The relations (\ref{L1L0})--(\ref{L1L1+}) show that the symmetry
algebra is nonlinear (quadratic). Using the commutation relations
one can write the right hand sides of quadratic products of the
operators (\ref{L1L0})--(\ref{L1L1+}) in various equivalent forms.
Each such a form of writing the algebra can, in principle, lead
to different BRST charges. It means, to control of constructing the
BRST operator we should fix an ordering for quadratic products of
operators in the algebra. We consider a most general ordering
prescription. All possible ways to order the operators in right hand
sides of (\ref{L1L0})--(\ref{L1L1+}) are described in terms of
arbitrary real parameters $\xi_1$, $\xi_2$, $\xi_3$, $\xi_4$,
$\xi_5$. The ordered commutation relations look like
\begin{eqnarray}
[L_1,\tilde{L}_0]&=&
\gamma r L_1
+(2-\xi_1)\beta rL_1^+L_2
+(2+\xi_1)\beta rL_2L_1^+
-4\beta rl_1^{\prime+}L_2
-4\beta rl_2'L_1^+
\nonumber
\\
&&{}
+(1-\xi_2)\beta r G_0L_1
+(1+\xi_2)\beta r L_1G_0
-2\beta rl_1'G_0
-2\beta rg_0'L_1
\nonumber
\\
&&{}
+(\xi_1-\xi_2)\beta r L_1
,
\label{L1L0xi}
\\
{}
[\tilde{L}_0,L_1^+]
&=&
\gamma r L_1^+
+(2+\xi_3)\beta rL_2^+L_1
+(2-\xi_3)\beta rL_1L_2^+
-4\beta rl_2^{\prime+}L_1
-4\beta rl_1'L_2^+
\nonumber
\\
&&{}
+(1+\xi_4)\beta r L_1^+G_0
+(1-\xi_4)\beta r G_0L_1^+
-2\beta r l_1^{\prime+}G_0
-2\beta r g_0'L_1^+
\nonumber
\\
&&{}
+(\xi_4-\xi_3) \beta r L_1^+
,
\label{L0L1+xi}
\\
{}
[L_1,L_1^+]
&=&
\tilde{L}_0-\gamma rG_0
+4(2-\beta)r(l_2^{\prime+}L_2+l_2'L_2^+)
-2(2-\beta)rg_0'G_0
\nonumber
\\
&&{}
+(2-\beta)r
\Bigl(
G_0^2
-\xi_5G_0
-(2+\xi_5)L_2^+L_2
-(2-\xi_5)L_2L_2^+
\Bigr)
.
\label{L1L1+xi}
\end{eqnarray}
Before we begin a construction of the BRST operator ones point out
that the algebra of the enlarged operators in the case under
consideration is very similar to one considered in \cite{Sevrin}.
The only difference of our case from \cite{Sevrin} consists in the
symmetry property of constants $f_{ij}^{km}$ (\ref{alg-enl}). In
\cite{Sevrin} this quantity is symmetric in upper indices
$f_{ij}^{km}=f_{ij}^{mk}$, but in the present paper we leave
them
having arbitrary symmetry by means of introducing the arbitrary
parameters $\xi_i$. At $\xi_i=0$ we have the symmetrized product of
the operators in rhs of (\ref{L1L0xi})--(\ref{L1L1+xi}) and our
algebra will be a partial case of the algebra considered in
\cite{Sevrin}. It was proved in \cite{Sevrin} that the BRST operator
includes the terms of first, thirds and seventh order in ghosts.
Therefore one can expect in our case the analogous terms in BRST
operator plus some extra terms stipulated by difference in symmetry
of the constants $f_{ij}^{km}$.

Now let us turn to construction of BRST operator. Demanding that
BRST operator be Hermitian (with respect to the new scalar product
(\ref{sprod})) we find equations on $\xi_i$. These equations leave
us only one arbitrary coefficient which we denote $\xi$, the others
are expressed through it as follows
\begin{eqnarray}
\xi=\xi_1=\xi_2=\xi_3=\xi_4,
&\qquad&
\xi_5=0.
\end{eqnarray}
In this case the commutators (\ref{L1L0xi})--(\ref{L1L1+xi})take the
form
\begin{eqnarray}
[L_1,\tilde{L}_0]&=&
\gamma r L_1
+(2-\xi)\beta rL_1^+L_2
+(2+\xi)\beta rL_2L_1^+
-4\beta rl_1^{\prime+}L_2
-4\beta rl_2'L_1^+
\nonumber
\\
&&{}
+(1-\xi)\beta r G_0L_1
+(1+\xi)\beta r L_1G_0
-2\beta rl_1'G_0
-2\beta rg_0'L_1
,
\label{L1L0x}
\\
{}
[\tilde{L}_0,L_1^+]
&=&
\gamma r L_1^+
+(2+\xi)\beta rL_2^+L_1
+(2-\xi)\beta rL_1L_2^+
-4\beta rl_2^{\prime+}L_1
-4\beta rl_1'L_2^+
\nonumber
\\
&&{}
+(1+\xi)\beta r L_1^+G_0
+(1-\xi)\beta r G_0L_1^+
-2\beta r l_1^{\prime+}G_0
-2\beta r g_0'L_1^+
,
\label{L0L1+x}
\\
{}
[L_1,L_1^+]
&=&
\tilde{L}_0-\gamma rG_0
+4(2-\beta)r(l_2^{\prime+}L_2+l_2'L_2^+)
-2(2-\beta)rg_0'G_0
\nonumber
\\
&&{}
+(2-\beta)r
( G_0^2 -2L_2^+L_2 -2L_2L_2^+ )
.
\label{L1L1+x}
\end{eqnarray}

To find the BRST operator we write it as a Hermitian operator
polynomial of seventh degree in ghosts and demand its nilpotency.
Such a procedure leads to the following result for the  BRST
operator $Q'$
\begin{eqnarray}
Q'
&=&
\eta_0\tilde{L}_0+\eta_1^+L_1+\eta_1L_1^+
+\eta_2^+L_2+\eta_2L_2^++\eta_{G}G_0
-\eta_1^+\eta_1{\cal{}P}_0
-\eta_2^+\eta_2{\cal{}P}_G
\nonumber
\\&&
{}
+(\eta_G\eta_1^++\eta_2^+\eta_1){\cal{}P}_1
+(\eta_1\eta_G+\eta_1^+\eta_2){\cal{}P}_1^+
+2\eta_G\eta_2^+{\cal{}P}_2
+2\eta_2\eta_G{\cal{}P}_2^+
\nonumber
\\
&&{}
+\beta r \eta_1^+\eta_0
\Bigl[
4l_1^{\prime+}{\cal{}P}_2+4l_2'{\cal{}P}_1^+
+2g_0'{\cal{}P}_1+2l_1'{\cal{}P}_G
\Bigr]
\nonumber
\\
&&{}
+\beta r \eta_0\eta_1
\Bigl[
4l_1'{\cal{}P}_2^++4l_2^{\prime+}{\cal{}P}_1
+2g_0'{\cal{}P}_1^++2l_1^{\prime+}{\cal{}P}_G
\Bigr]
\nonumber
\\
&&\hspace{-2em}
{}
-\beta r\eta_1^+\eta_0
\Bigl[
(2-\xi)L_1^+{\cal{}P}_2
+(2+\xi)L_2{\cal{}P}_1^+
+(1-\xi)G_0{\cal{}P}_1
+(1+\xi)L_1{\cal{}P}_G
\Bigr]
\nonumber
\\
&&\hspace{-2em}
{}
-\beta r\eta_0\eta_1
\Bigl[
(2-\xi)L_1{\cal{}P}_2^+
+(2+\xi)L_2^+{\cal{}P}_1
+(1-\xi)G_0{\cal{}P}_1^+
+(1+\xi)L_1^+{\cal{}P}_G
\Bigr]
\nonumber
\\
&&\hspace{-2em}
{}
-(2-\beta)r\eta_1^+\eta_1
\Bigl[
G_0{\cal{}P}_G-2L_2^+{\cal{}P}_2-2L_2{\cal{}P}_2^+
-2g_0'{\cal{}P}_G+4l_2^{\prime+}{\cal{}P}_2+4l_2'{\cal{}P}_2^+
\Bigr]
\nonumber
\\
&&
{}
+
\gamma r\eta_1^+\eta_1{\cal{}P}_G
+
\gamma r
\eta_0
\Bigl[
\eta_1^+{\cal{}P}_1
-
\eta_1{\cal{}P}_1^+
+
2\eta_2^+{\cal{}P}_2
-
2\eta_2{\cal{}P}_2^+
\Bigr]
\nonumber
\\
&&
{}
+3\beta\xi r \eta_0
\Bigl[
\eta_1\eta_2{\cal{}P}_1^+{\cal{}P}_2^+
+
\eta_1^+\eta_2^+{\cal{}P}_1{\cal{}P}_2
+
\eta_2^+\eta_1{\cal{}P}_1^+{\cal{}P}_2
\nonumber
\\
&&
\qquad\qquad\qquad{}
+
\eta_1^+\eta_2{\cal{}P}_2^+{\cal{}P}_1
+
\eta_2^+\eta_1{\cal{}P}_G{\cal{}P}_1
+
\eta_1^+\eta_2{\cal{}P}_1^+{\cal{}P}_G
\Bigr]
.
\label{Q'}
\end{eqnarray}
It is interesting to compare this result for BRST operator with
the result of \cite{Sevrin}. We see that BRST operator (\ref{Q'}) unlike
to \cite{Sevrin} has no terms of the seventh degree in the ghosts
(three ghost momenta and four ghost coordinates). The matter is
that,
according to \cite{Sevrin} the presence (or the absence) of such terms
in BRST operator depends on the number of operators whose
commutators have quadratic rhs. To get the seventh degree terms we
need at least four such operators. In our case we get only three
operators whose commutators have quadratic rhs
(\ref{L1L0x})--(\ref{L1L1+x}). These commutators are of
$\tilde{L}_0$, $L_1$, $L_1^+$ operators, therefore we have only
three ghost coordinates $\eta_0$, $\eta_1^+$, $\eta_1$ whereas the
terms of seventh degree in ghosts demand at least four ghost
coordinates. New point in compare with \cite{Sevrin} is appearance
in BRST operator the terms of fifth degree in ghosts. Their origin
is another symmetry property of the constants $f_{ij}^{km}$ in the
case under consideration in comparison with \cite{Sevrin}. If we put
$\xi=0$ (what corresponds to the symmetric ordering of the operators
in rhs of (\ref{L1L0x})--(\ref{L1L1+x})) the BRST operator $Q'$
(\ref{Q'}) has no terms of the fifth degree in ghosts as it should
be in accordance with \cite{Sevrin}.

Further we will show that the arbitrariness in BRST operator
stipulated by the parameter $\xi$ is resulted in arbitrariness of
introducing the auxiliary fields in the Lagrangians and hence does
not affect on dynamics of the basic field (\ref{PhysState}).

\section{Construction of Lagrangians}\label{Lagr-constr}

To construct the Lagrangian we use the procedure developed in
\cite{0505092} for massive bosonic higher spin fields in the flat
space. According to this procedure we should extract dependence of
$Q'$ (\ref{Q'}) on the ghosts $\eta_G$, ${\cal{}P}_G$
\begin{eqnarray}
Q'
&=&
Q+
\eta_G(\sigma+h)
-\eta_2^+\eta_2{\cal{}P}_G
+2\beta r\eta_0(\eta_1l_1^{\prime+}-\eta_1^+l_1'){\cal{}P}_G
\nonumber
\\
&&{}
+(1+\xi)\beta r\eta_0(\eta_1^+L_1-\eta_1L_1^+){\cal{}P}_G
-(2-\beta)r\eta_1^+\eta_1(G_0-2g_0'){\cal{}P}_G
\nonumber
\\
&&
{}
+\gamma r\eta_1^+\eta_1{\cal{}P}_G
+3\beta\xi r \eta_0
(\eta_1^+\eta_2{\cal{}P}_1^+-\eta_2^+\eta_1{\cal{}P}_1)
{\cal{}P}_G,
\end{eqnarray}
where
\begin{eqnarray}
\sigma+h
\equiv
G_0+\eta_1^+{\cal{}P}_1-\eta_1{\cal{}P}_1^+
+2\eta_2^+{\cal{}P}_2-2\eta_2{\cal{}P}_2^+
,
\label{sigma}
\qquad
\qquad
[\sigma,Q]=0,
\end{eqnarray}
\begin{eqnarray}
Q
&=&
\eta_0\tilde{L}_0+\eta_1^+L_1+\eta_1L_1^+
+\eta_2^+L_2+\eta_2L_2^+
-\eta_1^+\eta_1{\cal{}P}_0
+\eta_2^+\eta_1{\cal{}P}_1
+\eta_1^+\eta_2{\cal{}P}_1^+
\nonumber
\\
&&{}
+\beta r \eta_1^+\eta_0
\Bigl[
4l_1^{\prime+}{\cal{}P}_2+4l_2'{\cal{}P}_1^+
+2g_0'{\cal{}P}_1
\Bigr]
+\beta r \eta_0\eta_1
\Bigl[
4l_1'{\cal{}P}_2^++4l_2^{\prime+}{\cal{}P}_1
+2g_0'{\cal{}P}_1^+
\Bigr]
\nonumber
\\
&&
{}
-\beta r\eta_1^+\eta_0
\Bigl[
(2-\xi)L_1^+{\cal{}P}_2
+(2+\xi)L_2{\cal{}P}_1^+
+(1-\xi)G_0{\cal{}P}_1
\Bigr]
\nonumber
\\
&&
{}
-\beta r\eta_0\eta_1
\Bigl[
(2-\xi)L_1{\cal{}P}_2^+
+(2+\xi)L_2^+{\cal{}P}_1
+(1-\xi)G_0{\cal{}P}_1^+
\Bigr]
\nonumber
\\
&&
{}
+
(2-\beta)r\eta_1^+\eta_1
\Bigl[
2L_2^+{\cal{}P}_2+2L_2{\cal{}P}_2^+
-4l_2^{\prime+}{\cal{}P}_2-4l_2'{\cal{}P}_2^+
\Bigr]
\nonumber
\\
&&
{}
+
\gamma r
\eta_0
\Bigl[
\eta_1^+{\cal{}P}_1
-
\eta_1{\cal{}P}_1^+
+
2\eta_2^+{\cal{}P}_2
-
2\eta_2{\cal{}P}_2^+
\Bigr]
\nonumber
\\
&&
{}
+3\beta\xi r \eta_0
\Bigl[
\eta_1\eta_2{\cal{}P}_1^+{\cal{}P}_2^+
+
\eta_1^+\eta_2^+{\cal{}P}_1{\cal{}P}_2
+
\eta_2^+\eta_1{\cal{}P}_1^+{\cal{}P}_2
+
\eta_1^+\eta_2{\cal{}P}_2^+{\cal{}P}_1
\Bigr]
.
\label{Q}
\end{eqnarray}

We consider that the ghost operators act on the vacuum state as
follows
\begin{equation}
 {\cal{}P}_0|0\rangle
={\cal{}P}_G|0\rangle
=\eta_1|0\rangle
={\cal{}P}_1|0\rangle
=\eta_2|0\rangle
={\cal{}P}_2|0\rangle
=0
\label{ghostvac}
\end{equation}
and suppose \cite{0505092} that the vectors and the gauge parameters
do not depend on the ghost $\eta_G$. As a result we have from the
equation defining the physical vectors $Q'|\Psi\rangle=0$, where
$|\Psi\rangle$ is a vector in the extended space (including all
ghosts  except $\eta_G$)
\begin{align}
\label{QPsi}
&
Q|\Psi\rangle=0,
&&
(\sigma+h)|\Psi\rangle=0,
&&
gh(|\Psi\rangle)=0,
\\
&
\delta|\Psi\rangle=Q|\Lambda\rangle,
&&
(\sigma+h)|\Lambda\rangle=0,
&&
gh(|\Lambda\rangle)=-1,
\label{QLambda}
\\
&
\delta|\Lambda\rangle=Q|\Omega\rangle,
&&
(\sigma+h)|\Omega\rangle=0,
&&
gh(|\Omega\rangle)=-2.
\label{QOmega}
\end{align}
Since we can not write a gauge parameter with ghost number $-3$, the
chain of gauge transformation is finite.

Let us discuss construction of Lagrangian for the field with a given
value of the spin $s$. The middle equation of (\ref{QPsi}) is
equation for defining the possible values of the arbitrary parameter
$h$. Ones can see that it takes the values $-h=s+\frac{d}{2}-3$ and
these values are connected with the spin of the field. Having fixed
a value of the spin  we define the parameter $h$. This value of $h$
is substituted into the other equations (\ref{QPsi})--(\ref{QOmega})
(including operator $Q$). Let us denote $Q_s$ the operator $Q$
(\ref{Q}) where we substituted $s+\frac{d}{2}-3$ instead of $-h$
\begin{eqnarray}
Q_s&=&Q\bigm|_{-h\to{}s+\frac{d}{2}-3}
\end{eqnarray}
and let us denote the eigenvectors of the operator $\sigma$
(\ref{sigma}) corresponding to the eigenvalue $s+\frac{d}{2}-3$
as $|\chi\rangle_s$
\begin{eqnarray}
\label{eigen}
\sigma|\chi\rangle_s
&=&
(s+{\textstyle\frac{d}{2}}-3)|\chi\rangle_s,
\qquad
\qquad
\qquad
Q_s^2|\chi\rangle_s\equiv0.
\end{eqnarray}
Thus for a given spin-$s$ we have from (\ref{QPsi})--(\ref{QOmega})
the equation of motion and the gauge transformation
\begin{eqnarray}
&&Q_s|\Psi\rangle_s=0,
\label{QPsis}
\\
&&
\delta|\Psi\rangle_s=Q_s|\Lambda\rangle_s,
\\
&&
\delta|\Lambda\rangle_s=Q_s|\Omega\rangle_s
\label{QOmegas}
\end{eqnarray}
and the middle equations of (\ref{QPsi})--(\ref{QOmega}) are
satisfied by construction.

Next step \cite{0505092} is to extract the Hermitian ghosts
$\eta_0$, ${\cal{}P}_0$ in equations (\ref{QPsis})--(\ref{QOmegas}).
We have\footnote{Here and further ones assume that we substituted
$s+\frac{d}{2}-3$ instead of $-h$ in all the operators $l_i'$.}
\begin{eqnarray}
Q_s
&=&
\Delta{}Q+\eta_0\tilde{\tilde{L}}_0-\eta_1^+\eta_1{\cal{}P}_0,
\end{eqnarray}
\begin{eqnarray}
\Delta{}Q
&=&
\eta_1^+L_1+\eta_1L_1^+
+\eta_2^+L_2+\eta_2L_2^+
+\eta_2^+\eta_1{\cal{}P}_1
+\eta_1^+\eta_2{\cal{}P}_1^+
\nonumber
\\
&&
{}
+
(2-\beta)r\eta_1^+\eta_1
\Bigl[
2L_2^+{\cal{}P}_2+2L_2{\cal{}P}_2^+
-4l_2^{\prime+}{\cal{}P}_2-4l_2'{\cal{}P}_2^+
\Bigr]
,
\end{eqnarray}
\begin{eqnarray}
\tilde{\tilde{L}}_0
&=&
\tilde{L}_0
-\beta r \eta_1^+
\Bigl[
4l_1^{\prime+}{\cal{}P}_2+4l_2'{\cal{}P}_1^+
+2g_0'{\cal{}P}_1
\Bigr]
+\beta r \eta_1
\Bigl[
4l_1'{\cal{}P}_2^++4l_2^{\prime+}{\cal{}P}_1
+2g_0'{\cal{}P}_1^+
\Bigr]
\nonumber
\\
&&
{}
+\beta r\eta_1^+
\Bigl[
(2-\xi)L_1^+{\cal{}P}_2
+(2+\xi)L_2{\cal{}P}_1^+
+(1-\xi)G_0{\cal{}P}_1
\Bigr]
\nonumber
\\
&&
{}
-\beta r\eta_1
\Bigl[
(2-\xi)L_1{\cal{}P}_2^+
+(2+\xi)L_2^+{\cal{}P}_1
+(1-\xi)G_0{\cal{}P}_1^+
\Bigr]
\nonumber
\\
&&
{}
+
\gamma r
\Bigl[
\eta_1^+{\cal{}P}_1
-
\eta_1{\cal{}P}_1^+
+
2\eta_2^+{\cal{}P}_2
-
2\eta_2{\cal{}P}_2^+
\Bigr]
\nonumber
\\
&&
{}
+3\beta\xi r
\Bigl[
\eta_1\eta_2{\cal{}P}_1^+{\cal{}P}_2^+
+
\eta_1^+\eta_2^+{\cal{}P}_1{\cal{}P}_2
+
\eta_2^+\eta_1{\cal{}P}_1^+{\cal{}P}_2
+
\eta_1^+\eta_2{\cal{}P}_2^+{\cal{}P}_1
\Bigr]
.
\end{eqnarray}
for the operator $Q_s$
and
\begin{eqnarray}
|\Psi\rangle&=&|S\rangle+\eta_0|A\rangle,
\label{Psi=SA}
\\
|\Lambda\rangle&=&|\Lambda_0\rangle+\eta_0|\Lambda_1\rangle,
\qquad
\qquad
\qquad
|\Omega\rangle=
|\Omega_0\rangle
\end{eqnarray}
for the gauge parameters.
As a result we have the equations of motion
and the gauge transformation
\begin{align}
\label{EofM}
&
\tilde{\tilde{L}}_0|S\rangle-\Delta{}Q|A\rangle=0,
&&
\Delta Q|S\rangle-\eta_1^+\eta_1|A\rangle=0,
\\
&
\delta|S\rangle
=
\Delta Q|\Lambda_0\rangle-\eta_1^+\eta_1|\Lambda_1\rangle
,
&&
\delta|A\rangle=\tilde{\tilde{L}}_0|\Lambda_0\rangle
-\Delta Q|\Lambda_1\rangle
,
\label{gaugetr}
\\
&
\delta|\Lambda_0\rangle=\Delta Q|\Omega\rangle
,
&&
\delta|\Lambda_1\rangle=\tilde{\tilde{L}}_0|\Omega\rangle
\label{gaugetr-}
.
\end{align}

It is easy to show that the equations of motion (\ref{EofM}) can be
derived from the following Lagrangian
\begin{eqnarray}
\label{Ls}
{\cal{}L}_s
&=&
\langle{}S|K\Bigl\{ \tilde{\tilde{L}}_0|S\rangle-\Delta{}Q|A\rangle\Bigr\}
+
\langle{}A|K\Bigl\{-\Delta{}Q|S\rangle+\eta_1^+\eta_1|A\rangle\Bigr\}
\end{eqnarray}
which can also be written in a more compact form
\begin{eqnarray}
\label{Lscompact}
{\cal{}L}_s&=&\int d\eta_0\,\,
{}_s\langle\Psi|KQ_s|\Psi\rangle_s.
\end{eqnarray}

Now we fix the arbitrary parameter $m_0^2$. It is defined from the
condition of reproducing the conditions (\ref{constr}) for the
basic vector $|\Phi\rangle$ (\ref{PhysState}). The general vector
$|\Psi\rangle$ includes the basic vector $|\Phi\rangle$ as follows
\begin{eqnarray}
|\Psi\rangle&=&|\Phi\rangle+  |\Phi_A\rangle
,
\end{eqnarray}
where the vector $|\Phi_A\rangle$ includes only the auxiliary fields
as its components. One can show, using the part of equations of
motion and gauge transformations, that the vector $|\Phi_A\rangle$
can be completely removed. The details are given in
Appendix~\ref{proof}. As a result we obtain the equation of motion
for the basic vector (\ref{QPsis}) in the form
\begin{eqnarray}
(l_0-m^2+M^2)|\Phi\rangle
=l_1|\Phi\rangle
=l_2|\Phi\rangle
=0,
\label{-m+M}
\end{eqnarray}
where $M^2$ is defined in (\ref{Mass}). We see that the Lagrangian
reproduces the basic conditions (\ref{constr}) if $M^2 = m^2$. It
leads to
\begin{eqnarray}
m_0^2&=&
-\alpha{\textstyle\frac{d(d-4)}{4}}\,r
-(\beta-2)h(h-2)r+\gamma hr
.
\label{m0}
\end{eqnarray}
It is intersting to note that in case of $\alpha=\gamma=0$,
$\beta=2$ we will have $m_0^2=0$. Such a situation was considered in
\cite{BRST-AdS} where the higher spin massless bosonic fields in AdS
space were studied.

One can prove that the Lagrangian (\ref{Lscompact}) indeed
reproduces the basic conditions (\ref{constr}). The details of
the proof
are given in Appendix~\ref{proof}. Relation (\ref{Lscompact}) where
the parameter $m_0^2$ is fixed by (\ref{m0}) is our final result.

Construction of the Lagrangian describing propagation of all massive
bosonic fields in AdS space simultaneously is analogous to that in
the flat space \cite{0505092} and we do not consider it here.

Let us turn to examples.

\section{Examples}\label{examples}

\subsection{Spin-0 field}

For a scalar field with spin $s=0$ we have $h=3-\frac{d}{2}$. Then
the vector $|\Psi\rangle_s$ which satisfies the condition
(\ref{eigen}) and has the proper ghost number (\ref{QPsi}) looks
like
\begin{eqnarray}
|\Psi\rangle_0&=&\varphi(x)|0\rangle
.
\end{eqnarray}
Then using (\ref{m0}), (\ref{K}) the relation for Lagrangian
(\ref{Ls}) gives
\begin{eqnarray}
{\cal{}L}_0
&=&
\varphi\Bigl\{
\nabla^2
+r(6-2d)
+m^2
\Bigr\}\varphi
\end{eqnarray}
It is easy to see that this Lagrangian reproduces the equation
(\ref{Eq-0}) for $s=0$.

\subsection{Spin-1 field}

For a vector field we have $h=2-\frac{d}{2}$. Then the vector
$|\Psi\rangle_s$ and the gauge parameter $|\Lambda\rangle_s$, which
satisfy the condition (\ref{eigen}) and have the proper ghost
numbers (\ref{QPsi}), (\ref{QLambda}), look like
\begin{eqnarray}
|\Psi\rangle_1&=& \bigl[-ia^{\mu+}A_\mu(x)+b_1^+A(x)\bigr]|0\rangle
+\eta_0{\cal{}P}_1^+\varphi(x)|0\rangle,
\\
|\Lambda\rangle_1&=&
{\cal{}P}_1^+\lambda(x)|0\rangle
\end{eqnarray}
Using (\ref{m0}), (\ref{K}) we find the Lagrangian
(\ref{Ls}) for vector field
\begin{eqnarray}
{\cal{}L}&=&
-A^\mu
\Bigl[
\bigl(\nabla^2-r(d-1)+m^2\bigr)A_\mu-\nabla_\mu\varphi
\Bigr]
\nonumber
\\
&&{}
+
\frac{m^2}{m_1^2}\,
A
\Bigl[
\bigl(\nabla^2-4r+m^2\bigr)A-m_1\varphi
\Bigr]
+
\varphi\Bigl[\varphi-\nabla^\mu{}A_\mu
-\frac{m^2}{m_1}
\,A
\Bigr]
\label{L1'}
\end{eqnarray}
and the gauge transformations (\ref{gaugetr})
\begin{eqnarray}
\delta{}A_\mu(x)=\nabla_\mu\lambda,
\qquad
\delta{}A(x)=m_1\lambda,
\qquad
\delta\varphi(x)=(\nabla^2+m^2)\lambda.
\label{gaugetr1}
\end{eqnarray}
The relations (\ref{L1'}) and (\ref{gaugetr1}) are the final result
for massive vector filed. We obtain the gauge invariant theory in
terms of vector filed $A_\mu$ and two auxiliary scalar fields $A$
and $\varphi$.

Let us consider the massless limit of this theory. To do that we put
$m\to0$ in (\ref{L1'})\footnote{Ones remind that arbitrary parameter
$m_1$ arose in Section 5 when we constructed the additional parts of
the operators. In general it has no relation to physical mass $m$ and should
not tend to zero in massless limit.}. Then the field $A$ vanishes
and we obtain the theory of the massless field $s=1$ in AdS space.

Now we show how the Lagrangian (\ref{L1'}) is transformed to the
conventional form. Let us rescale the field $A$
\begin{eqnarray}
\frac{m}{m_1}A&=&A' \to A.
\end{eqnarray}
In this case Lagrangian (\ref{L1'}) and the gauge transformation
(\ref{gaugetr1}) are rewritten as
\begin{eqnarray}
{\cal{}L}&=&
-A^\mu
\Bigl\{
\bigl[\nabla^2+m^2-r(d-1)\bigr]
A_\mu
-\nabla_\mu\varphi
\Bigr\}
\nonumber
\\
&&{}
+
A
\Bigl\{
\bigl[\nabla^2+m^2\bigr]
A
-m\varphi
\Bigr\}
+
\varphi\Bigl\{\varphi-\nabla^\mu{}A_\mu-mA
\Bigr\}
,
\label{LagrM1}
\end{eqnarray}
\begin{eqnarray}
\delta{}A_\mu(x)=\nabla_\mu\lambda,
\qquad
\delta{}A(x)=m\lambda,
\qquad
\delta\varphi(x)=\bigl(\nabla^2+m^2\bigr)\lambda
\end{eqnarray}
and they become independent of $m_1$.

Ones exclude the field $\varphi(x)$ from Lagrangian (\ref{LagrM1})
with the help of its equation of motion. It leads to
\begin{eqnarray}
{\cal{}L}&=&
{\textstyle\frac{1}{2}}\,F^{\mu\nu}F_{\mu\nu}
-
(mA_\mu-\nabla_\mu{}A)(mA^\mu-\nabla^\mu{}A)
,
\end{eqnarray}
where
\begin{eqnarray}
F_{\mu\nu}&=&\nabla_\mu{}A_\nu-\nabla_\nu{}A_\mu
.
\end{eqnarray}
Now removing the field $A$ with the help of its gauge
transformation we arrive at the conventional form of Lagrangian for
spin-1 field (up to an overall factor)
\begin{eqnarray}
{\cal{}L}&=&
{\textstyle\frac{1}{2}}\,F^{\mu\nu}F_{\mu\nu}
-
m^2A_\mu A^\mu
.
\end{eqnarray}

\subsection{Spin-2 field}
For spin-2 field we have $h=1-\frac{d}{2}$. Then the vector
$|\Psi\rangle_s$ and the gauge parameter $|\Lambda\rangle_s$, which
satisfy the condition (\ref{eigen}) and have the proper ghost
numbers (\ref{QPsi}), (\ref{QLambda}), look like
\begin{eqnarray}
|S_1\rangle&=& \Bigr[
{\textstyle\frac{(-i)^2}{2}}\,a^{\mu+}a^{\nu+}H_{\mu\nu}(x)
+b_2^+H_1(x) -ia^{\mu+}b_1^+A_\mu(x) +b_1^{+2}\varphi(x)
\Bigl]|0\rangle
\\
|S_2\rangle
&=&
H(x)|0\rangle,
\\
|A_1\rangle&=& \Bigl[ -ia^{\mu+}H_\mu(x)+b_1^+A(x) \Bigr]|0\rangle,
\hspace{7em} |A_2\rangle=H_2(x)|0\rangle,
\\
|\lambda_1\rangle&=& \Bigl[ -ia^{\mu+}\lambda_\mu(x)+b_1^+\lambda(x)
\Bigr]|0\rangle, \hspace{7.4em} |\lambda_2\rangle=
\lambda_2(x)|0\rangle.
\end{eqnarray}
Here $|S_i\rangle$, $|A_i\rangle$, $|\lambda_i\rangle$ are defined
in (\ref{S})--(\ref{Lambda0}).

Using (\ref{m0}), (\ref{Ls}), (\ref{K}) we find Lagrangian for the
massive spin-2 field in the form
\begin{eqnarray}
{\cal{}L}
&=&
{\textstyle\frac{1}{2}}\,H^{\mu\nu}
\Bigl\{
\Bigl[\nabla^2+m^2-2r\Bigr]H_{\mu\nu}
-2(\beta-1)rg_{\mu\nu}H^\sigma{}_\sigma
\nonumber
\\
&&\hspace{7em}
{}
+(2+\xi)\beta rg_{\mu\nu}H
+g_{\mu\nu}H_2
-2\nabla_\mu H_\nu
\Bigr\}
\nonumber
\\
&&
{}
-
\Bigl(\frac{d-2}{2}\,H_1+\frac{m^2}{m_1^2}\,\varphi\Bigr)
\Bigl\{
\Bigl[\nabla^2+m^2+2r(d-1)\Bigr]H_1
\nonumber
\\
&&\hspace{5em}
{}
-4\beta{}r\Bigl[\frac{d-2}{2}\,H_1+\frac{m^2}{m_1^2}\,\varphi\Bigr]
+\beta(2-\xi)rH-H_2
\Bigr\}
\nonumber
\\
&&
{}
-
\frac{m^2}{m_1^2}\,A^\mu
\Bigl\{
\Bigl[\nabla^2+m^2+r(d-1)\Bigr]A_\mu
-\nabla_\mu A-m_1H_\mu
\Bigr\}
\nonumber
\\
&&\hspace{-3em}
{}
+
\frac{m^2}{m_1^2}
\Bigl[
\frac{2m^2+2r(d-1)}{m_1^2}\,\varphi
-
H_1
\Bigr]
\Bigl\{
\Bigl(\nabla^2+m^2+2r(d-1)\Bigr)\varphi
-m_1A
\Bigr\}
\nonumber
\\
&&
\hspace{-3em}
{}
-H
\Bigl\{
\Bigl(
\nabla^2+m^2+2r(d-1)+2\xi\beta r
\Bigr)
H
-\nabla^\mu{}H_\mu
-\frac{m^2}{m_1}\,A
+H_2
\nonumber
\\
&&\hspace{3em}
{}
-{\textstyle\frac{1}{2}}\,(2+\xi)\beta r H^\mu{}_\mu
+\beta{}r(2-\xi)\,\frac{d-2}{2}\,H_1
+\beta{}r(2-\xi)\,\frac{m^2}{m_1^2}\,\varphi
\Bigr\}
\nonumber
\\
&&{}
+
H^\mu\Bigl\{
\nabla^\nu H_{\mu\nu}-H_\mu-\nabla_\mu H
+\frac{m^2}{m_1}\,A_\mu
\Bigr\}
\nonumber
\\
&&{}
-
\frac{m^2}{m_1^2}\,A
\Bigl\{
\nabla^\mu A_\mu-m_1(H+H_1)-A
+\frac{2m^2+2r(d-1)}{m_1}\,\varphi
\Bigr\}
\nonumber
\\
&&{}
+
H_2
\Bigl\{
{\textstyle\frac{1}{2}}\,H^\mu{}_\mu
+{\textstyle\frac{d-2}{2}}\,H_1
+\frac{m^2}{m_1^2}\,\varphi-H
\Bigr\}
\label{L-2}
.
\end{eqnarray}
Using (\ref{gaugetr}), (\ref{m0}) ones find the gauge
transformations
\begin{eqnarray}
&&
\delta{}H_{\mu\nu}
=
\nabla_\mu\lambda_\nu+\nabla_\nu\lambda_\mu
-g_{\mu\nu}\lambda_2
,
\hspace{4em}
\delta{}H_1=\lambda_2,
\\
&&
\delta{}A_\mu=\nabla_\mu\lambda+m_1\lambda_\mu,
\hspace{8.5em}
\delta\varphi=m_1\lambda,
\\
&&
\delta{}H=
\nabla^\mu\lambda_\mu+\frac{m^2}{m_1}\,\lambda
-\lambda_2,
\\
&&
\delta{}H_\mu
=
\Bigl(\nabla^2+m^2+r(d-1)\Bigr)\lambda_\mu,
\\
&&
\delta{}A
=
\Bigl(\nabla^2+m^2+2r(d-1)\Bigr)\lambda,
\\
&&
\delta{}H_2
=
(2-\xi)\beta r\nabla^\mu\lambda_\mu
-(2+\xi)\beta r\,\frac{m^2}{m_1}\,\lambda
\nonumber
\\
&&\hspace{3em}
{}
+(\nabla^2+m^2+2r(d-1)+\beta r(2-2d+\xi))\lambda_2
.
\end{eqnarray}
The relation (\ref{L-2}) and the above gauge transformations are our
final result for massive spin-2 theory in AdS space. We get the
gauge theory in terms of basic filed $H_{\mu\nu}$ and some number of
auxiliary fields.

In massless limit $m\to0$\footnote{We assume as in Subsection 8.2
that the arbitrary parameter $m_1$ does not tend to zero in this
limit.} in (\ref{L-2}) we get Lagrangian for the massless spin-2
field in terms of the fields $H_{\mu\nu}$, $H_\mu$, $H$, $H_1$ and
$H_2$.

Now we rewrite the Lagrangian (\ref{L-2}) in  more conventional
form. Let us make the following  transformation. First, we redefine
the fields
\begin{eqnarray}
\frac{d-2}{2}\,H_1+\frac{m^2}{m_1^2}\varphi
=
\frac{d-2}{2}\,H_1^{\,\prime}
&\to&
\frac{d-2}{2}\,H_1
,
\\
\frac{m}{m_1}\,A=A'\to A,
\qquad
\frac{m}{m_1}\,A_\mu=A_\mu'\to A_\mu,
&\quad&
\frac{m^2}{m_1^2}\,\varphi
=
\varphi'\to\varphi
,
\end{eqnarray}
and the gauge parameters
\begin{eqnarray}
\frac{m}{m_1}\,\lambda=\lambda'\to\lambda.
\end{eqnarray}
Then we use the equations of motion for $H_\mu$, $A$ in (\ref{L-2})
and remove field $H_1$ with the help of its gauge transformation.
After that the gauge parameters $\lambda$ and $\lambda_2$ are not
independent
\begin{eqnarray}
\lambda_2+\frac{2m}{d-2}\lambda
&=&0.
\end{eqnarray}
Next we use the equations of motion for the fields $H_2$ and $H$ in
previously obtained Lagrangian. Finally, after one more rescaling
the field $\varphi$
\begin{eqnarray}
\Bigl[2(d-1)\Bigl(\frac{1}{d-2}+\frac{r}{m^2}\Bigr)\Bigr]^{1/2}
\,\varphi=\varphi'\to\varphi
\end{eqnarray}
we arrive at Zinoviev's Lagrangian \cite{Zinoviev}
\begin{eqnarray}
{\cal{}L}
&=&
{\textstyle\frac{1}{2}}\,H^{\mu\nu}
\Bigl[\nabla^2+m^2-2r\Bigr]H_{\mu\nu}
-
{\textstyle\frac{1}{2}}\,H^\mu{}_\mu
\Bigl[\nabla^2+m^2+r(d-3)\Bigr]H^\nu{}_\nu
\nonumber
\\
&&\hspace{5em}
+
H^{\mu\nu}\Bigl[
\nabla_\mu\nabla_\nu{}H^\sigma_\sigma
-\nabla_\mu\nabla^\sigma H_{\sigma\nu}
\Bigr]
\nonumber
\\
&&
{}
-
A^\mu
\Bigl\{
\Bigl[\nabla^2+r(d-1)\Bigr]A_\mu
-\nabla_\mu\nabla^\nu A_\nu
\Bigr\}
+
\varphi
\Bigl(\nabla^2-\frac{d}{d-2}\,m^2\Bigr)\varphi
\nonumber
\\
&&
{}
+
\Bigl[2(d-1)\Bigl(\frac{m^2}{d-2}+r\Bigr)\Bigr]^{1/2}
\varphi
\Bigl[
mH^\mu_\mu
-2\nabla^\mu A_\mu
\Bigr]
\nonumber
\\
&&
{}
+2m\Bigl[H^\sigma_\sigma\nabla^\mu{}A_\mu-H^{\mu\nu}\nabla_\mu{}A_\nu\Bigr]
\label{L-2new}
\end{eqnarray}
with the gauge transformations
\begin{eqnarray}
\delta{}H_{\mu\nu}
&=&
\nabla_\mu\lambda_\nu+\nabla_\nu\lambda_\mu
+g_{\mu\nu}\frac{2m}{d-2}\lambda
,
\\
\delta{}A_\mu
&=&
\nabla_\mu\lambda+m\lambda_\mu,
\\
\delta\varphi&=&
\Bigl[2(d-1)\Bigl(\frac{m^2}{d-2}+r\Bigr)\Bigr]^{1/2}
\lambda
.
\end{eqnarray}
It is easy to see that at
\begin{eqnarray}
r+\frac{m^2}{d-2}=0
\end{eqnarray}
Lagrangian (\ref{L-2new}) splits into a sum of two independent parts
with helicities $\pm2$, $\pm1$ ($H_{\mu\nu}$, $A_\mu$) and the
scalar field $\varphi$. Thus the field $H_{\mu\nu}$ becomes partial
massless at
\begin{eqnarray}
m^2&=&-(d-2)r
\end{eqnarray}
in accordance with \cite{Deser}.

\section{Summary}\label{summary}
We have developed the gauge invariant approach to deriving the
Lagrangians for bosonic massive higher spin fields in arbitrary
dimensional AdS space. The Lagrangian includes the Stueckelberg
fields, providing the gauge invariance in the massive theory, and the
complete set of the auxiliary fields which should be introduced for
Lagrangian formulation in higher spin theory. The approach is based
on BRST construction for special nonlinear symmetry algebra
formulated in the paper.

We begin with imbedding the massive higher spin fields into the
vectors of auxiliary Fock space. All such fields are treated as the
components of the vectors of the Fock space and the theory is
completely formulated in terms of such vectors. The conditions
defining the irreducible representation of the AdS group with given
mass and spin are realized by differential operators acting in this
Fock space. The above conditions are interpreted as the constraints
on the vectors of the Fock space and generate the closed nonlinear
symmetry algebra which is the main object of our analysis. As we proved,
derivation of the correct Lagrangian demands an extension and
deformation of the algebra. BRST construction for such nonlinear
symmetry algebra is given. It is shown that the BRST operator
corresponding to the extended deformed algebra defines the
consistent Lagrangian dynamics for bosonic massive fields of any
value of spin. We constructed the gauge invariant Lagrangians in
terms of Fock space in the concise form for any higher spin
fields propagating in AdS space of arbitrary dimension. It is
interesting to point out that the gauge transformations in the theory
under consideration are reducible, the corresponding order of
reducibility is equal to unit. As an example of the general scheme we
obtained the gauge invariant Lagrangians and the corresponding gauge
transformations for the component  spin-0, spin-1, and spin-2
massive fields in the explicit form. The Lagrangians for component
massive fields with other spins can be analogously found using the
simple enough manipulations with creation and annihilation operators
in the Fock space.

The main results of the paper are given by the relations
(\ref{Ls}), (\ref{Lscompact})
where Lagrangian for the massive field with arbitrary integer spin is
constructed,
and
(\ref{gaugetr}), (\ref{gaugetr-})
where the gauge transformations for the fields and the gauge
transformations for the gauge parameters are written down.

The procedure for Lagrangian construction developed here for higher
spin massive bosonic fields can also be applied to fermionic higher
spin theories in AdS background and for fields with mixed symmetry
tensor or tensor-spinor fields (see  \cite{0101201} where bosonic
massless fields with mixed symmetry were considered in Minkowski
space). The results obtained open a principle possibility for
derivation of the interacting vertices for massive higher spin
fields in AdS space.

\section*{Acknowledgements}
The authors are grateful to R.R. Metsaev, H. Takata, M. Tsulaia,
M.A. Vasiliev for discussions. The work was partially supported by
the INTAS grant, project INTAS-03-51-6346, the RFBR grant, project
No.\ 06-02-16346 and grant for LRSS, project No.\ 4489.2006.2. Work
of I.L.B and P.M.L was supported in part by the DFG grant, project
No.\ 436 RUS 113/669/0-3 and joint RFBR-DFG grant, project No.\
06-02-04012.

\appendix
\section*{Appendix}

\section{Differential calculus in auxiliary Fock space.}\label{Fock-Ap}
\renewcommand{\theequation}{\Alph{section}.\arabic{equation}}
\setcounter{equation}{0}

We consider an arbitrary $d$-dimensional Riemann space with  metric
$g_{\mu\nu}(x)$ which can be expressed in terms of the vielbein
$e^b_{\mu}(x)$
\begin{eqnarray}
\nonumber
&&g_{\mu\nu}(x)=\eta_{bc}\;e^b_{\mu}(x)\;e^c_{\nu}(x)\;,\quad
g_{\mu\lambda}(x)g^{\lambda\nu}(x)=\delta_{\mu}^{\nu}\;,
\\
\label{F1}
\\
\nonumber && e^a_{\mu}(x)\;e^{\mu}_b(x)=\delta^a_b\;,\quad
e^{\mu}_b(x)=g^{\mu\nu}(x)\;\eta_{bc}\;e^c_{\nu}(x)\;,\quad
\eta_{bc}\eta^{cd}=\delta_b^d\;,
\end{eqnarray}
where $\eta_{bc}$ is the Minkowski metric with the signature
$\eta_{bc}=(+,-,...,-)$.

We introduce the creation and annihilation operators $a^+_b$ and
$a_b$ carrying flat-space indices $b$
\begin{eqnarray}
&&[a_b,a^+_c]=-\eta_{bc}\;,\quad [a_b,a_c]=0\;,\quad
[a^+_b,a^+_c]=0\;,
\\
&& a_b|0\rangle=0 \;,
\qquad\quad \langle
0|0\rangle=1 \;, \quad \partial_{\mu} a^b=\partial_{\mu}
a^{b+}=0\;,
\end{eqnarray}
where  $|0\rangle$ is the vacuum vector of the Fock space. Any
$s$-particle vector $|\Phi^s\rangle$ can be presented in the form
\begin{eqnarray}\label{F3}
|\Phi^s\rangle=\Phi_{b_1...b_s}(x)\;
a^{b_1+}...a^{b_s+}|0\rangle\;,\quad
N|\Phi^s\rangle=s|\Phi^s\rangle\;,\quad N=-a^{b+} a_b\;,\quad
a^{b+}=\eta^{bc}a_b^+\;,
\end{eqnarray}
where $\Phi_{b_1...b_s}(x)$ is a symmetric tensor field of rank $s$.
It is useful to introduce the creation and annihilation operators
$a^+_{\mu}(x)$ and $a_{\mu}(x)$ by the relations
\begin{eqnarray}\label{F4}
a^+_{\mu}(x)=e^b_{\mu}(x)\;a^+_b\;,\quad
a_{\mu}(x)=e^b_{\mu}(x)\;a_b\;.
\end{eqnarray}
They obey the properties
\begin{eqnarray}\label{F5}
[a^+_{\mu}(x),a^+_{\nu}(x)]=0\;,\quad
[a_{\mu}(x),a_{\nu}(x)]=0\;,\quad
[a_{\mu}(x),a^+_{\nu}(x)]=g_{\mu\nu}(x)\;,\quad
a_{\mu}(x)|0\rangle=0\;
\end{eqnarray}
and create the s-particle vector
\begin{eqnarray}\label{F6}
|\Phi^s\rangle&=&\Phi_{\mu_1...\mu_s}(x)\;
a^{\mu_1+}(x)...a^{\mu_s+}(x)|0\rangle\;,\quad
N|\Phi^s\rangle=s|\Phi^s\rangle\;,\\
\nonumber
 &&N=-a^{\mu+}(x)a_{\mu}(x)\;,\quad
a^{\mu+}=g^{\mu\nu}(x)a^+_{\nu}(x)\;.
\end{eqnarray}
where $\Phi_{\mu_1...\mu_s}(x)$ is a symmetric tensor on the Riemann
space.

Ones introduce the operator ${\cal D}_{\mu}$ acting on any operator
tensor with curved-space and with flat-space indices by the rule
\begin{eqnarray}\nonumber
\label{D1}
{\cal
D}_{\mu}Q^{\cdots}_{\cdots}&=&
(\nabla_\mu Q^{\cdots}_{\cdots})
- \omega_{\mu\;\;\;c}^{\;\;\;b}\;[a^{c+}a_b,Q^{\cdots}_{\cdots}]\;,
\end{eqnarray}
where the dots denote the curved-space and  flat-space indices. The
$\nabla_{\mu}$ is the proper covariant derivative.  It is clear that
this operator possesses the Leibniz rule
\begin{eqnarray}\label{D2}
{\cal D}_{\mu}\Big(A\cdot B\Big)=\Big({\cal D}_{\mu}A\Big)\cdot
B+A\cdot\Big({\cal D}_{\mu}B\Big).
\end{eqnarray}

From definition of ${\cal D}_{\mu}$ it follows that it acts on usual
tensor fields (which do not depend on creation and annihilation
operators) as the covariant derivative
\begin{eqnarray}\label{D6}
{\cal
D}_{\mu}\Phi_{\mu_1\mu_2...\mu_s}(x)=\triangledown_{\mu}\Phi_{\mu_1\mu_2...\mu_s}(x)\;,\quad
{\cal
D}_{\mu}\Phi_{b_1b_2...b_s}(x)=\triangledown_{\mu}\Phi_{b_1b_2...b_s}(x).
\end{eqnarray}
In particular we have
\begin{eqnarray}
{\cal
D}_{\mu}g_{\alpha\beta}(x)=\triangledown_{\mu}g_{\alpha\beta}(x)=0.
\end{eqnarray}
Consider the action of ${\cal D}_{\mu}$ on the creation and
annihilation operators with flat-space indices. Ones obtain
\begin{eqnarray}\label{D7}
{\cal D}_{\mu}a^b=\partial_{\mu}a^b+\omega_{\mu\;\;\;c}^{\;\;b}a^c-
\omega_{\mu\;\;\;d}^{\;\;c}\;[a^{d+}a_c,a^b]=
(\omega_{\mu\;\;c}^{\;\;b}+\omega_{\mu c}^{\;\;\;\;b})a^c=0\;,\quad
{\cal D}_{\mu}a^{b+}=0,
\end{eqnarray}
because of $\partial_{\mu} a^b=\partial_{\mu} a^{b+}=0$ and
$\omega_{\mu\;\;c}^{\;\;b}=-\omega_{\mu c}^{\;\;\;\;b}$. Therefore
\begin{eqnarray}\nonumber\label{D8}
{\cal D}_{\mu}a^{\nu}&=&\partial_{\mu}e^{\nu}_b
a^b+\Gamma^{\nu}_{\;\;\alpha\mu}e^{\alpha}_b a^b-
\omega_{\mu\;\;\;c}^{\;\;d}e^{\nu}_d\;[a^{c+}a_b,a^d]=
(\partial_{\mu}e^{\nu}_b+\Gamma^{\nu}_{\;\;\alpha\mu}e^{\alpha}_b+\omega_{\mu
b}^{\;\;\;\;c}e^{\nu}_c)a^b=\\
\label{D9}
&=&(\nabla_{\mu}e^{\nu}_b)a^b=0\;,\quad
{\cal D}_{\mu}a^{\nu+}=0.
\end{eqnarray}

The property of covariant constancy of the creation and annihilation
operators (\ref{D7}), (\ref{D9}) allow us to consider the operator
${\cal D}_{\mu}$ as representation of the covariant derivative on
vectors $|\Phi^s\rangle$ if we assume that ${\cal
D}_{\mu}|0\rangle=0$. Indeed, we have
\begin{eqnarray}
{\cal
D}_{\mu}|\Phi^s\rangle=\Big(\triangledown_{\mu}\Phi_{\mu_1\mu_2...\mu_s}(x)\Big)
a^{\mu_1+}(x)...a^{\mu_s+}(x)|0\rangle=
\Big(\triangledown_{\mu}\Phi_{b_1b_2...b_s}(x)\Big)a^{b_1+}...a^{b_s+}|0\rangle
\end{eqnarray}

In a similar way we can find the action of the commutator of the
covariant derivative on the vectors (\ref{F6})
\begin{eqnarray}\label{F19}
[{\cal D}_{\mu},{\cal D}_{\nu}]|\Phi^s\rangle=
-R^{\alpha}_{\;\;\beta\mu\nu}a^{\beta+}a_{\alpha}|\Phi^s\rangle=
-{\cal R}^b_{\;\;c\mu\nu}a^{c+}a_b|\Phi^s\rangle\;,
\end{eqnarray}
where we used notations
\begin{eqnarray}
\label{F20}
R^{\alpha}_{\;\;\beta\mu\nu}&=&\partial_{\mu}\Gamma^{\alpha}_{\;\;\beta\nu}-
\partial_{\nu}\Gamma^{\alpha}_{\;\;\beta\mu}-
\Gamma^{\lambda}_{\;\;\beta\mu}\Gamma^{\alpha}_{\;\;\lambda\nu}+
\Gamma^{\lambda}_{\beta\nu}\Gamma^{\alpha}_{\;\;\lambda\mu}\;,
\\
\label{F23}
{\cal R}^b_{\;\;c\mu\nu}&=&\partial_{\mu}\omega_{\nu\;\;
c}^{\;\;b}-
\partial_{\nu}\omega_{\mu\;\;
c}^{\;\;b}- \omega_{\mu\;\; d}^{\;\;b}\;\omega_{\nu\;\; c}^{\;\;d}+
\omega_{\nu\;\; d}^{\;\;b}\;\omega_{\mu\;\; c}^{\;\;d}
\end{eqnarray}
for the curvature tensors.

\section{Calculation of the additional parts}\label{Ap-AddParts}
\renewcommand{\theequation}{\Alph{section}.\arabic{equation}}
\setcounter{equation}{0}

In this Appendix we show how the representation of the algebra given
in Table~\ref{table'} can be constructed in terms of some creation
and annihilation operators.

Let us consider a representation of this algebra with the vector
$|0\rangle_V$ annihilated by the operators $l_1'$ and
$l_2'$
\begin{eqnarray}
l_1'|0\rangle_V=l_2'|0\rangle_V=0
\label{0V}
\end{eqnarray}
and being the eigenvector of the operators $\tilde{l}_0'$,
$g_0'$ and $l'$
\begin{eqnarray}
\tilde{l}_0'|0\rangle_V=m_0^2|0\rangle_V,
\qquad
g_0'|0\rangle_V=h|0\rangle_V,
\qquad
l'|0\rangle_V=m_2^2|0\rangle_V,
\label{mh}
\end{eqnarray}
where $m_0$, $m_2$ are arbitrary constants with dimension of
mass and $h$ is a dimensionless arbitrary constant\footnote{The
representation given by (\ref{0V}) and (\ref{mh}) is called in
the mathematical literature the Verma module. It explains the
subscript $V$ at the vectors.}.
They are the arbitrary constants which must be contained in the
additional parts of Hermitian operators.
Next we choose the basis vectors as follows
\begin{eqnarray}
\label{basisV}
&&
|n_1,n_2\rangle_V
=
\left(\frac{l^{\prime+}}{m_1}\right)^{n_1}
(l_2^{\prime+})^{n_2}|0\rangle_V
,
\end{eqnarray}
where $m_1$ is an arbitrary nonzero constant with dimension of
mass. It may be constructed from the parameters of the theory
$m_1=f(m,r)\neq0$.
It is easy to find how the operators $l'$, $l_1^{\prime+}$,
$l_2^{\prime+}$, $g_0'$ act on these basis vectors
\begin{eqnarray}
\label{B4}
&&
l'|n_1,n_2\rangle_V=m_2^2|n_1,n_2\rangle_V,
\\
&&
l_1^{\prime+}|n_1,n_2\rangle_V=m_1|n_1+1,n_2\rangle_V,
\\
&&
l_2^{\prime+}|n_1,n_2\rangle_V=|n_1,n_2+1\rangle_V,
\\
&&
g_0'|n_1,n_2\rangle_V=(n_1+2n_2+h)|n_1,n_2\rangle_V.
\label{B7}
\end{eqnarray}

Calculation of how the rest operators $\tilde{l}_0'$, $l_1'$,
$l_2'$ acts on the basis vectors is much more difficult.
First we find that
\begin{eqnarray}
\label{B8}
\tilde{l}_0'|n_1,n_2\rangle_V
&=&
2\gamma rn_2|n_1,n_2\rangle_V
+
(l_2^{\prime+})^{n_2}\tilde{l}_0'|n_1,0\rangle_V
,
\\
\label{B9}
l_1'|n_1,n_2\rangle_V
&=&
-m_1n_2|n_1+1,n_2-1\rangle_V
+
(l_2^{\prime+})^{n_2}l_1|n_1,0\rangle_V
,
\\
l_2'|n_1,n_2\rangle_V
&=&
(n_2-1+n_1+h)n_2|n_1,n_2-1\rangle_V
+
(l_2^{\prime+})^{n_2}l_2|n_1,0\rangle_V
\label{B10}
.
\end{eqnarray}
Thus it remains to calculate how the operators $\tilde{l}_0'$,
$l_1'$, $l_2'$ act on the basis vectors $|n_1,0\rangle_V$.

To do this we define some auxiliary operators and their action on
the vector $|0\rangle_V$
\begin{align}
&
K_0\equiv g_0^{\prime2}-2g_0'-4l_2^{\prime+}l_2'
,
&&
K_0|0\rangle_V=h(h-2)|0\rangle_V
,
\\
&
K_1\equiv
\Bigl[K_0,l_1^{\prime+}\Bigr]
=
4l_2^{\prime+}l_1'
+2l_1^{\prime+}g_0'-l_1^{\prime+}
,
&&
K_1|0\rangle_V=m_1(2h-1)|1,0\rangle_V
.
\end{align}
\begin{eqnarray}
K_2&\equiv&
\Bigl[K_1,l_1^{\prime+}\Bigr]
=
2l_1^{\prime+2}+4l_2^{\prime+}K_2'
,
\\
K_2'&=&
\tilde{l}_0'
+m^2+\alpha{\textstyle\frac{d(d-4)}{4}}\,r
+(\beta-2)rK_0
-\gamma rg_0',
\\
K_2|0\rangle_V
&=&
4M^2|0,1\rangle_V
+
2m_1^2|2,0\rangle_V
,
\qquad
K_2'|0\rangle_V
=
M^2|0\rangle_V,
\end{eqnarray}
where
\begin{eqnarray}
M^2&=&
m^2+m_0^2
+\alpha{\textstyle\frac{d(d-4)}{4}}\,r
+(\beta-2)rh(h-2)-\gamma rh.
\end{eqnarray}
In terms of these operators we have
\begin{eqnarray}
\Bigl[l_1',l_1^{\prime+}\Bigr]
=
K_2'
,
\qquad
\Bigl[\tilde{l}_0',l_1^{\prime+}\Bigr]
=
-\beta rK_1+\gamma rl_1^{\prime+}
,
\qquad
\Bigl[K_2,l_1^{\prime+}\Bigr]=
-8l_2^{\prime+}rK_1.
\end{eqnarray}
Using the above formulas we obtain
\begin{eqnarray}
\tilde{l}_0'|n_1,0\rangle_V
&=&
m_0|n_1,0\rangle_V
+\gamma rn_1 |n_1,0\rangle_V
\nonumber
\\
&&
{}
-
\beta r (2h-1)
\sum_{k=0}
\left[\frac{-8r}{m_1^2}\right]^k
C_{2k+1}^{n_1}|n_1-2k,k\rangle_V
\nonumber
\\
&&
{}
-
2\beta r
\sum_{k=0}
\left[\frac{-8r}{m_1^2}\right]^k
C_{2k+2}^{n_1}|n_1-2k,k\rangle_V
\nonumber
\\
&&
{}
-
\frac{4\beta{}rM^2}{m_1^2}
\sum_{k=0}
\left[\frac{-8r}{m_1^2}\right]^k
C_{2k+2}^{n_1}|n_1-2k-2,k+1\rangle_V
\label{B18}
\end{eqnarray}
where $C^n_k=\displaystyle{\frac{n!}{k!(n-k)!}}\;$.
Substituting (\ref{B18}) into (\ref{B8}) ones get
\begin{eqnarray}
\tilde{l}_0'|n_1,n_2\rangle_V
&=&
m_0|n_1,n_2\rangle_V
+\gamma r(n_1+2n_2) |n_1,n_2\rangle_V
\nonumber
\\
&&
{}
-
\beta r (2h-1)
\sum_{k=0}
\left[\frac{-8r}{m_1^2}\right]^k
C_{2k+1}^{n_1}|n_1-2k,n_2+k\rangle_V
\nonumber
\\
&&
{}
-
2\beta r
\sum_{k=0}
\left[\frac{-8r}{m_1^2}\right]^k
C_{2k+2}^{n_1}|n_1-2k,n_2+k\rangle_V
\nonumber
\\
&&
{}
-
\frac{4\beta{}rM^2}{m_1^2}
\sum_{k=0}
\left[\frac{-8r}{m_1^2}\right]^k
C_{2k+2}^{n_1}|n_1-2k-2,n_2+k+1\rangle_V
.
\label{B19}
\end{eqnarray}

Analogously one can first find
\begin{eqnarray}
l_1'|n_1,0\rangle_V
&=&
m_1
\frac{2h-1}{4}
\sum_{k=1}
\left[\frac{-8r}{m_1^2}\right]^k
C_{2k}^{n_1}
|n_1-2k+1,k-1\rangle_V
\nonumber
\\
&&
{}
+
\frac{m_1}{2}
\sum_{k=1}
\left[\frac{-8r}{m_1^2}\right]^k
C_{2k+1}^{n_1}
|n_1-2k+1,k-1\rangle_V
\nonumber
\\
&&
{}
+
\frac{M^2}{m_1}
\sum_{k=0}
\left[\frac{-8r}{m_1^2}\right]^k
C_{2k+1}^{n_1}
|n_1-2k-1,k\rangle_V
,
\label{B20}
\end{eqnarray}
\begin{eqnarray}
l_2'|n_1,0\rangle_V
&=&
\frac{1-2h}{4}
\sum_{k=1}
\left[\frac{-8r}{m_1^2}\right]^k
C_{2k+1}^{n_1}
|n_1-2k,k-1\rangle_V
\nonumber
\\
&&
{}
-\frac{1}{2}
\sum_{k=1}
\left[\frac{-8r}{m_1^2}\right]^k
C_{2k+2}^{n_1}
|n_1-2k,k-1\rangle_V
\nonumber
\\
&&
{}
-
\frac{M^2}{m_1^2}
\sum_{k=0}
\left[\frac{-8r}{m_1^2}\right]^k
C_{2k+2}^{n_1}
|n_1-2k-2,k\rangle_V
\label{B21}
\end{eqnarray}
and then substituting (\ref{B20}) and (\ref{B21}) into (\ref{B9})
and (\ref{B10}) respectively we find how operators $l_1'$ and
$l_2'$ act on the basis vectors $|n_1,n_2\rangle_V$
\begin{eqnarray}
l_1'|n_1,n_2\rangle_V
&=&
-m_1n_2|n_1+1,n_2-1\rangle_V
+
\nonumber
\\
&&
{}
+
m_1
\frac{2h-1}{4}
\sum_{k=1}
\left[\frac{-8r}{m_1^2}\right]^k
C_{2k}^{n_1}
|n_1-2k+1,n_2+k-1\rangle_V
\nonumber
\\
&&
{}
+
\frac{m_1}{2}
\sum_{k=1}
\left[\frac{-8r}{m_1^2}\right]^k
C_{2k+1}^{n_1}
|n_1-2k+1,n_2+k-1\rangle_V
\nonumber
\\
&&
{}
+
\frac{M^2}{m_1}
\sum_{k=0}
\left[\frac{-8r}{m_1^2}\right]^k
C_{2k+1}^{n_1}
|n_1-2k-1,n_2+k\rangle_V
,
\label{B22}
\end{eqnarray}
\begin{eqnarray}
l_2'|n_1,n_2\rangle_V
&=&
(n_2-1+n_1+h)n_2|n_1,n_2-1\rangle_V
\nonumber
\\
&&{}
-
\frac{2h-1}{4}
\sum_{k=1}
\left[\frac{-8r}{m_1^2}\right]^k
C_{2k+1}^{n_1}
|n_1-2k,n_2+k-1\rangle_V
\nonumber
\\
&&
{}
-\frac{1}{2}
\sum_{k=1}
\left[\frac{-8r}{m_1^2}\right]^k
C_{2k+2}^{n_1}
|n_1-2k,n_2+k-1\rangle_V
\nonumber
\\
&&
{}
-
\frac{M^2}{m_1^2}
\sum_{k=0}
\left[\frac{-8r}{m_1^2}\right]^k
C_{2k+2}^{n_1}
|n_1-2k-2,n_2+k\rangle_V
.
\label{B23}
\end{eqnarray}

Now let us turn to construction of a representation of the
operator algebra given in Table~\ref{table'} in terms of
creation and annihilation operators.
The number of pairs of these operators and their statistics is
defined by the number of operators used in the definition of the
basis vectors (\ref{basisV}).
Thus we introduce two pairs of the bosonic creation and
annihilation operators with the standard commutation relations
\begin{eqnarray}
[b_1,b_1^+]=1,
\qquad
[b_2,b_2^+]=1,
\end{eqnarray}
corresponding to $l_1'$, $l_1^{\prime+}$ and $l_2'$,
$l_2^{\prime+}$ respectively.
After this we map the basis vectors (\ref{basisV}) and the basis
vectors of the Fock space generated by $b_1^+$, $b_2^+$
\begin{eqnarray}
|n_1,n_2\rangle_V&\longleftrightarrow&
(b_1^+)^{n_1}(b_2^+)^{n_2}|0\rangle=|n_1,n_2\rangle
\end{eqnarray}
and find from (\ref{B4})--(\ref{B7}), (\ref{B19}), (\ref{B22}),
(\ref{B23}) form of the operators in terms of the creation and
annihilation operators
\begin{align}
&
l'=m_2^2,
&&
l_1^{\prime+}
=
m_1b_1^+,
\\
&
l_2^{\prime+}
=
b_2^+,
&&
g_0'=
b_1^+b_1+2b_2^+b_2+h,
\end{align}
\begin{eqnarray}
\tilde{l}_0' &=& m_0^2+\gamma r(b_1^+b_1+2b_2^+b_2) - \beta r
(2h-1)\,b_1^+ \sum_{k=0}^{\infty} \left[\frac{-8r}{m_1^2}\right]^k
\frac{b_2^{+k}\,b_1^{2k+1}}{(2k+1)!} \nonumber
\\
&& {} - 2\beta r \, b_1^{+2} \sum_{k=0}^{\infty}
\left[\frac{-8r}{m_1^2}\right]^k
\frac{b_2^{+k}\,b_1^{2k+2}}{(2k+2)!} + \frac{1}{2}\beta M^2
\sum_{k=1}^{\infty} \left[\frac{-8r}{m_1^2}\right]^k
\frac{b_2^{+k}\,b_1^{2k}}{(2k)!}
,
\end{eqnarray}
\begin{eqnarray}
l_1'
&=&
-m_1b_1^+b_2
+
m_1b_1^+
\frac{2h-1}{4}
\sum_{k=1}^{\infty}
\left[\frac{-8r}{m_1^2}\right]^k
\frac{(b_2^+)^{k-1}\,b_1^{2k}}{(2k)!}
\nonumber
\\
&& {} + \frac{1}{2} \, m_1b_1^{+2} \sum_{k=1}^{\infty}
\left[\frac{-8r}{m_1^2}\right]^k
\frac{(b_2^+)^{k-1}\,b_1^{2k+1}}{(2k+1)!} + \frac{M^2}{m_1}
\sum_{k=0}^{\infty} \left[\frac{-8r}{m_1^2}\right]^k
\frac{b_2^{+k}\,b_1^{2k+1}}{(2k+1)!}
,
\end{eqnarray}
\begin{eqnarray}
l_2'
&=&
(b_2^+b_2+b_1^+b_1+h)b_2
-
\frac{2h-1}{4}
\,
b_1^+
\sum_{k=1}^{\infty}
\left[\frac{-8r}{m_1^2}\right]^k
\frac{(b_2^+)^{k-1}\,b_1^{2k+1}}{(2k+1)!}
\nonumber
\\
&& {} -\frac{1}{2} \, b_1^{+2} \sum_{k=1}^{\infty}
\left[\frac{-8r}{m_1^2}\right]^k
\frac{(b_2^+)^{k-1}\,b_1^{2k+2}}{(2k+2)!} - \frac{M^2}{m_1^2}
\sum_{k=0}^{\infty} \left[\frac{-8r}{m_1^2}\right]^k
\frac{b_2^{+k}\,b_1^{2k+2}}{(2k+2)!}.
\end{eqnarray}
Finally we put the arbitrary constant $m_2^2$ to be equal to
$m_2^2=-m^2-\alpha{\textstyle\frac{d(d-4)}{4}}\,r$.
Thus we have obtained the expressions of the additional parts
(\ref{l0'})--(\ref{M'})
for the operator algebra given in Table~\ref{table'}.

Let us find an explicit expression for the operator
$K$ which used in the definition of the scalar product
(\ref{sprod}).
The defining relations for this operator are given by
(\ref{H0})--(\ref{H2}).
These relations shall be satisfied for any $|\Psi\rangle$ if they
shall be satisfied for the basis vectors of the Fock space
$|n_1,n_2\rangle$.
Then using arguments of \cite{0101201,0410215}
one can check that the following operator satisfy (\ref{H0})--(\ref{H2})
\begin{eqnarray}
K=Z^+Z,
\qquad
Z=\sum_{n_1,n_2=0}^{\infty}
|n_1,n_2\rangle_V\frac{1}{n_1!n_2!}\langle n_1,n_2|\;.
\end{eqnarray}
For practical calculations it is useful to note that
${}_{V}\langle{}n_1',n_2'|n_1,n_2\rangle_V\sim\delta^{n_1+2n_2}_{n_1'+2n_2'}$.
For low numbers $n_1+2n_2$ the operator $K$ is
\begin{eqnarray}
K&=&|0\rangle\langle0|
+
\frac{M^2}{m_1^2}
b_1^+|0\rangle
\langle0|\,b_1
+
hb_2^+|0\rangle
\langle0|\,b_2
-
\frac{M^2}{2m_1^2}
\Bigl(
b_1^{+2}|0\rangle\langle0|\,b_2
+
b_2^+|0\rangle\langle0|\,b_1^2
\Bigr)
\nonumber
\\
&&
{}
+
\frac{M^2}{m_1^2}
\frac{M^2+r(1-2h)}{2m_1^2}\;
b_1^{+2}|0\rangle
\langle0|\,b_1^2
+
\ldots
\label{K}
\end{eqnarray}
This expression for the operator $K$ is used when we give the examples.

\section{Removing of the auxiliary fields}\label{proof}
\renewcommand{\theequation}{\Alph{section}.\arabic{equation}}
\setcounter{equation}{0}

In this Appendix we explain how the conditions (\ref{constr}) on
basic vector (\ref{PhysState}) can be obtained from equations of
motion (\ref{EofM}) following from Lagrangian (\ref{Lscompact}).

First we explicitly extract the dependence of the fields and the
gauge parameters on the ghost fields
\begin{eqnarray}
|S\rangle&=&
|S_1\rangle
+\eta_1^+{\cal{}P}_1^+|S_2\rangle
+\eta_1^+{\cal{}P}_2^+|S_3\rangle
+\eta_2^+{\cal{}P}_1^+|S_4\rangle
+\eta_2^+{\cal{}P}_2^+|S_5\rangle
\nonumber
\\
&&{}
+\eta_1^+\eta_2^+{\cal{}P}_1^+{\cal{}P}_2^+|S_6\rangle,
\label{S}
\\
|A\rangle&=&
{\cal{}P}_1^+|A_1\rangle
+{\cal{}P}_2^+|A_2\rangle
+\eta_1^+{\cal{}P}_1^+{\cal{}P}_2^+|A_3\rangle
+\eta_2^+{\cal{}P}_1^+{\cal{}P}_2^+|A_4\rangle
\label{A}
\\
|\Lambda\rangle&=&|\Lambda_0\rangle+\eta_0|\Lambda_1\rangle,
\label{Lambda}
\\
|\Lambda_0\rangle&=&
{\cal{}P}_1^+|\lambda_1\rangle
+{\cal{}P}_2^+|\lambda_2\rangle
+\eta_1^+{\cal{}P}_1^+{\cal{}P}_2^+|\lambda_3\rangle
+\eta_2^+{\cal{}P}_1^+{\cal{}P}_2^+|\lambda_4\rangle
\label{Lambda0}
\\
|\Lambda_1\rangle&=&
{\cal{}P}_1^+{\cal{}P}_2^+|\lambda_5\rangle,
\label{Lambda1}
\\
|\Omega\rangle&=&
{\cal{}P}_1^+{\cal{}P}_2^+|\omega\rangle.
\label{Omega}
\end{eqnarray}

In this notation the Lagrangian (\ref{Lscompact}) takes the form
\begin{eqnarray}
{\cal{}L}&=&
\langle{}S_1|K
\bigl\{
\tilde{L}_0|S_1\rangle-L_1^+|A_1\rangle
-L_2^+|A_2\rangle
-\beta{}r\Bigl[(2+\xi)L_2^+-4l_2^{\prime+}\Bigr]|S_2\rangle
\nonumber
\\
&&\qquad{}
-\beta{}r\Bigl[(1-\xi)G_0-2g_0'\Bigr]|S_1\rangle
-3\gamma{}r|S_1\rangle
-3r\beta\xi|S_1\rangle
\bigr\}
\nonumber
\\
&&{}
-
\langle{}S_2|K
\bigl\{
\tilde{L}_0|S_2\rangle-L_1|A_1\rangle+|A_2\rangle
 -L_2^+|A_3\rangle
\nonumber
\\
&&\qquad{}
+\beta{}r\Bigl[(1-\xi)G_0-2g_0'\Bigr]|S_2\rangle
+\beta{}r\Bigl[(2+\xi)L_2-4l_2'\Bigr]|S_1\rangle
\nonumber
\\
&&\qquad{}
+\beta{}r\Bigl[(2-\xi)L_1^+-4l_1^{\prime+}\Bigr]|S_4\rangle
-\gamma{}r|S_2\rangle
+3r\beta\xi|S_2\rangle
\bigr\}
\nonumber
\\
&&
-
\langle{}S_4|K
\bigl\{
\tilde{L}_0|S_3\rangle-L_1|A_2\rangle+L_1^+|A_3\rangle
\nonumber
\\
&&\qquad{}
+\beta{}r\Bigl[(2-\xi)L_1-4l_1'\Bigr]|S_2\rangle
+\beta{}r\Bigl[(2-\xi)L_1^+-4l_1^{\prime+}\Bigr]|S_5\rangle
\nonumber
\\
&&\qquad{}
+(2-\beta)r\Bigl[2L_2-4l_2'\Bigr]|A_1\rangle
-(2-\beta)r\Bigl[2L_2^+-4l_2^{\prime+}\Bigr]|A_4\rangle
\bigr\}
\nonumber
\\
&&
-
\langle{}S_3|K
\bigl\{
\tilde{L}_0|S_4\rangle-L_2|A_1\rangle
-L_2^+|A_4\rangle
\bigr\}
\nonumber
\\
&&
-
\langle{}S_5|K
\bigl\{
\tilde{L}_0|S_5\rangle-L_2|A_2\rangle
-|A_3\rangle+L_1^+|A_4\rangle
\nonumber
\\
&&\qquad{}
+\beta{}r\Bigl[(2-\xi)L_1-4l_1'\Bigr]|S_4\rangle
-\beta{}r\Bigl[(1-\xi)G_0-2g_0'\Bigr]|S_5\rangle
\nonumber
\\
&&\qquad{}
+\beta{}r\Bigl[(2+\xi)L_2^+-4l_2^{\prime+}\Bigr]|S_6\rangle
+\gamma{}r|S_5\rangle
+3r\beta\xi|S_5\rangle
\bigr\}
\nonumber
\\
&&
+
\langle{}S_6|K
\bigl\{
\tilde{L}_0|S_6\rangle+L_2|A_3\rangle-L_1|A_4\rangle
+\beta{}r\Bigl[(1-\xi)G_0-2g_0'\Bigr]|S_6\rangle
\nonumber
\\
&&\qquad{}
-\beta{}r\Bigl[(2+\xi)L_2-4l_2'\Bigr]|S_5\rangle
+3\gamma{}r|S_6\rangle
-3r\beta\xi|S_6\rangle
\bigr\}
\nonumber
\\
&&
-
\langle{}A_1|K
\bigl\{
L_1|S_1\rangle-L_1^+|S_2\rangle-L_2^+|S_3\rangle
  -|A_1\rangle
\nonumber
\\
&&\qquad{}
+(2-\beta)r\Bigl[2L_2^+-4l_2^{\prime+}\Bigr]|S_4\rangle
\bigr\}
\nonumber
\\
&&
-
\langle{}A_2|K
\bigl\{
L_2|S_1\rangle+|S_2\rangle
 -L_2^+|S_5\rangle-L_1^+|S_4\rangle
\bigr\}
\nonumber
\\
&&
+
\langle{}A_3|K
\bigl\{
L_2|S_2\rangle+|S_5\rangle
-L_1|S_4\rangle+L_2^+|S_6\rangle
\bigr\}
\nonumber
\\
&&
-
\langle{}A_4|K
\bigl\{
L_1|S_5\rangle-L_2|S_3\rangle
+L_1^+|S_6\rangle+|A_4\rangle
\nonumber
\\
&&\qquad{}
-(2-\beta)r\Bigl[2L_2-4l_2^{\prime}\Bigr]|S_4\rangle
\bigr\}
.
\label{Lagr}
\end{eqnarray}
Here we see at $\beta=\gamma=0$ (these values don't correspond
to that which give $l_0$ (\ref{l0-kaz}))
many terms in (\ref{Lagr}) disappear and the expression for the
Lagrangian is simplified.

The equations of motion which follow from Lagrangian (\ref{Lagr}) are
\begin{eqnarray}
\label{S1}
&&
\tilde{L}_0|S_1\rangle-L_1^+|A_1\rangle-L_2^+|A_2\rangle
-\beta{}r\Bigl[(2+\xi)L_2^+-4l_2^{\prime+}\Bigr]|S_2\rangle
\nonumber
\\
&&\qquad{}
-\beta{}r\Bigl[(1-\xi)G_0-2g_0'+3\xi\Bigr]|S_1\rangle
-3\gamma{}r|S_1\rangle
=0,
\\
\label{S2}
&&
\tilde{L}_0|S_2\rangle-L_1|A_1\rangle+|A_2\rangle
 -L_2^+|A_3\rangle
+\beta{}r\Bigl[(1-\xi)G_0-2g_0'+3\xi\Bigr]|S_2\rangle
-\gamma{}r|S_2\rangle
\nonumber
\\
&&\qquad{}
+\beta{}r\Bigl[(2+\xi)L_2-4l_2'\Bigr]|S_1\rangle
+\beta{}r\Bigl[(2-\xi)L_1^+-4l_1^{\prime+}\Bigr]|S_4\rangle
 =0,
\\
\label{S3}
&&
\tilde{L}_0|S_3\rangle-L_1|A_2\rangle+L_1^+|A_3\rangle
+\beta{}r\Bigl[(2-\xi)L_1-4l_1'\Bigr]|S_2\rangle
+\beta{}r\Bigl[(2-\xi)L_1^+-4l_1^{\prime+}\Bigr]|S_5\rangle
\nonumber
\\
&&\qquad{}
+(2-\beta)r\Bigl[2L_2-4l_2'\Bigr]|A_1\rangle
-(2-\beta)r\Bigl[2L_2^+-4l_2^{\prime+}\Bigr]|A_4\rangle
=0,
\\
\label{S4}
&&
\tilde{L}_0|S_4\rangle-L_2|A_1\rangle
-L_2^+|A_4\rangle
=0,
\\
\label{S5}
&&
\tilde{L}_0|S_5\rangle-L_2|A_2\rangle
-|A_3\rangle+L_1^+|A_4\rangle
-\beta{}r\Bigl[(1-\xi)G_0-2g_0'-3\xi\Bigr]|S_5\rangle
+\gamma{}r|S_5\rangle
\nonumber
\\
&&\qquad{}
+\beta{}r\Bigl[(2-\xi)L_1-4l_1'\Bigr]|S_4\rangle
+\beta{}r\Bigl[(2+\xi)L_2^+-4l_2^{\prime+}\Bigr]|S_6\rangle
=0,
\\
\label{S6}
&&
\tilde{L}_0|S_6\rangle+L_2|A_3\rangle
-L_1|A_4\rangle
+\beta{}r\Bigl[(1-\xi)G_0-2g_0'-3\xi\Bigr]|S_6\rangle
+3\gamma{}r|S_6\rangle
\nonumber
\\
&&\qquad{}
-\beta{}r\Bigl[(2+\xi)L_2-4l_2'\Bigr]|S_5\rangle
=0,
\\
\label{A1}
&&
L_1|S_1\rangle-L_1^+|S_2\rangle
-L_2^+|S_3\rangle-|A_1\rangle
+(2-\beta)r\Bigl[2L_2^+-4l_2^{\prime+}\Bigr]|S_4\rangle
=0,
\\
\label{A2}
&&
L_2|S_1\rangle+|S_2\rangle
 -L_2^+|S_5\rangle-L_1^+|S_4\rangle=0,
\\
\label{A3}
&&
L_2|S_2\rangle+|S_5\rangle
-L_1|S_4\rangle+L_2^+|S_6\rangle=0,
\\
\label{A4}
&&
L_1|S_5\rangle-L_2|S_3\rangle+L_1^+|S_6\rangle
  +|A_4\rangle
-(2-\beta)r\Bigl[2L_2-4l_2^{\prime}\Bigr]|S_4\rangle
=0.
\end{eqnarray}

These equations of motion and the Lagrangian (\ref{Lagr}) are
invariant under the gauge transformations
\begin{eqnarray}
&&\delta|S_1\rangle=
  L_1^+|\lambda_1\rangle
  +L_2^+|\lambda_2\rangle,
\label{dS1}
\\
&&\delta|S_2\rangle=
  L_1|\lambda_1\rangle
  -|\lambda_2\rangle
  +L_2^+|\lambda_3\rangle,
\\
&&\delta|S_3\rangle=
    L_1|\lambda_2\rangle
    -L_1^+|\lambda_3\rangle
    -|\lambda_5\rangle
-(2-\beta)r\Bigl[2L_2-4l_2^{\prime}\Bigr]|\lambda_1\rangle
\nonumber
\\
&&
\qquad\qquad
{}
+(2-\beta)r\Bigl[2L_2^+-4l_2^{\prime+}\Bigr]|\lambda_4\rangle
,
\\
&&\delta|S_4\rangle=
   L_2|\lambda_1\rangle
   +L_2^+|\lambda_4\rangle
\\
&&\delta|S_5\rangle=
  L_2|\lambda_2\rangle
  +|\lambda_3\rangle
  -L_1^+|\lambda_4\rangle
\\
&&\delta|S_6\rangle=
  -L_2|\lambda_3\rangle
  +L_1|\lambda_4\rangle
\\
&&\delta|A_1\rangle=
  (\tilde{L}_0-2\gamma{}r)|\lambda_1\rangle
  +L_2^+|\lambda_5\rangle,
\\
&&\delta|A_2\rangle=
  (\tilde{L}_0-\gamma{}r)|\lambda_2\rangle
  -L_1^+|\lambda_5\rangle
  +\beta{}r\Bigl[(2-\xi)L_1-4l_1'\Bigr]|\lambda_1\rangle
\nonumber
\\
&&\qquad\qquad{}
  -\beta{}r\Bigl[(1-\xi)G_0-2g_0'\Bigr]|\lambda_2\rangle
  -\beta{}r\Bigl[(2+\xi)L_2^+-4l_2^{\prime+}\Bigr]|\lambda_3\rangle
,
\\
&&\delta|A_3\rangle=
  (\tilde{L}_0+\gamma{}r)|\lambda_3\rangle
  -L_1|\lambda_5\rangle
  +\beta{}r\Bigl[(2+\xi)L_2-4l_2'\Bigr]|\lambda_2\rangle
\nonumber
\\
&&
\qquad\qquad
{}
+\beta{}r\Bigl[(1-\xi)G_0-2g_0'\Bigr]|\lambda_3\rangle
+\beta{}r\Bigl[(2-\xi)L_1^+-4l_1^{\prime+}\Bigr]|\lambda_4\rangle
,
\\
&&
\delta|A_4\rangle=
  (\tilde{L}_0+2\gamma{}r)|\lambda_4\rangle-L_2|\lambda_5\rangle
,
\label{dA4}
\end{eqnarray}
in its turn the gauge parameters are not uniquely defined but are
invariant under transformations
\begin{align}
&\delta|\lambda_1\rangle=
   -L_2^+|\omega\rangle,
&&
\delta|\lambda_2\rangle=
   L_1^+|\omega\rangle,
&&
\delta|\lambda_3\rangle=
   L_1|\omega\rangle,
\\
&\delta|\lambda_4\rangle=
  L_2|\omega\rangle,
&&
\delta|\lambda_5\rangle=
  \tilde{L}_0|\omega\rangle.
\end{align}
One can see again that the gauge transformations
(\ref{dS1})--(\ref{dA4}) are simplified at $\beta=\gamma=0$.

\subsection*{C 1. The gauge fixing}

First we discard the gauge parameter $|\lambda_5\rangle$ and consider
$|\lambda_1\rangle$, $|\lambda_2\rangle$, $|\lambda_3\rangle$,
$|\lambda_4\rangle$ to be independent.

Next
we remove dependence of $|S_1\rangle$ on the operator
$b_1^+$ using the parameter $|\lambda_1\rangle$ and
using the gauge parameter $|\lambda_2\rangle$ we can
remove the field $|S_3\rangle$ using some components of
$|\lambda_2\rangle$
\begin{eqnarray}
\delta|S_3\rangle
&=&
L_1|\lambda_2\rangle+\ldots
=
\Bigl(\frac{M^2}{m_1}b_1+\ldots
\Bigr)
|\lambda_2\rangle
+
\ldots
\end{eqnarray}
After this we have used parameter $|\lambda_1\rangle$ completely
and we have the restricted parameter $|\lambda_2\rangle$:
$b_1|\lambda_2\rangle=0$, i.e. $|\lambda_2\rangle$ is
independent of $b_1^+$.
Using this restricted gauge parameter $|\lambda_2\rangle$ we can
remove dependence of $|S_1\rangle$ on
$b_2^+$
getting
$|S_1\rangle=|\Phi\rangle$.
After this the parameter $|\lambda_2\rangle$ is exhausted.
Next we remove dependence of the vectors
$|S_2\rangle$ and $|S_4\rangle$
on $b_2^+$
with the help of the parameters $|\lambda_3\rangle$ and
$|\lambda_4\rangle$
respectively.
Now we have used all the gauge parameters.
Let us show that the rest auxiliary fields are zero as
consequence of the equations of motion in this
gauge.

\subsection*{C 2. Removing of the auxiliary fields with the equations of motion}
After the gauge fixing we have the gauge conditions
\begin{eqnarray}
|S_1\rangle=|\Phi\rangle,
\qquad
b_2|S_2\rangle=0,
\qquad
|S_3\rangle=0,
\qquad
b_2|S_4\rangle=0,
\label{thegauge}
\end{eqnarray}
where $|\Phi\rangle$ is the basic vector (\ref{PhysState}).
Let us look at the equation of motion (\ref{A2})
\begin{eqnarray}
l_2|\Phi\rangle+|S_2\rangle-L_2^+|S_5\rangle-L_1^+|S_4\rangle=0.
\end{eqnarray}
Acting on this equation by the operator $b_2$ and using
(\ref{thegauge}) we get
\begin{eqnarray}
b_2L_2^+|S_5\rangle=0
&\Longrightarrow&
|S_5\rangle=0
.
\end{eqnarray}

Acting by $b_2$ on (\ref{A1}) we get
\begin{eqnarray}
b_2|A_1\rangle=2r(\beta-2)|S_4\rangle
&\Longrightarrow&
b_2b_2|A_1\rangle=0.
\label{a1-s4}
\end{eqnarray}
Thus we can decompose $|A_1\rangle$
\begin{eqnarray}
|A_1\rangle=|A_1'\rangle+b_2^+|A_1''\rangle,
\end{eqnarray}
and (\ref{a1-s4}) takes the form
\begin{eqnarray}
|A_1''\rangle=2r(\beta-2)|S_4\rangle.
\end{eqnarray}

Acting twice by $b_2$ on (\ref{S1}) we get
\begin{eqnarray}
b_2b_2L_2^+|A_2\rangle=0
&\Longrightarrow&
b_2|A_2\rangle=0,
\end{eqnarray}
thus $|A_2\rangle$ does not depend on $b_2^+$.

Let us expand the fields
$|S_2\rangle$,
$|S_4\rangle$,
$|A_1'\rangle$,
$|A_1''\rangle$,
$|A_2\rangle$,
in powers of $b_1^+$
\begin{align}
&
|S_2\rangle=\sum_{k=0}^{s-2}(b_1^+)^k|S_{2k}\rangle,
&&
|S_4\rangle=\sum_{k=0}^{s-3}(b_1^+)^k|S_{4k}\rangle,
\\
&
|A_1'\rangle=\sum_{k=0}^{s-1}(b_1^+)^k|A_{1k}'\rangle,
&&
|A_1''\rangle=\sum_{k=0}^{s-3}(b_1^+)^k|A_{1k}''\rangle,
&&
|A_2\rangle=\sum_{k=0}^{s-2}(b_1^+)^k|A_{2k}\rangle
.
\end{align}
Then decomposing the equations of motion in powers of $b_1^+$
and $b_2^+$ ones obtain (here we assume (\ref{m0}))
\subsubsection*{Equations of motion (\ref{S1})}
\begin{subequations}
\label{S1-1}
\begin{eqnarray}
(b_1^+)^s
&&
-m_1|A_{1,s-1}'\rangle=0
,
\\
(b_1^+)^{s-1}
&&
-m_1|A_{1,s-2}'\rangle
=
l_1^+|A_{1,s-1}'\rangle
,
\\
(b_1^+)^k
&&
-m_1|A_{1,k-1}'\rangle
=
l_1^+|A_{1k}'\rangle
+l_2^+|A_{2k}\rangle
+\beta{}r(2+\xi)l_2^+|S_{2k}\rangle
,
\\
(b_1^+)^0
&&
\Bigl(l_0+4(2-\beta)l_2^+l_2\Bigr)|\Phi\rangle
=
l_1^+|A_{10}'\rangle
+l_2^+|A_{20}\rangle
+\beta{}r(2+\xi)l_2^+|S_{20}\rangle
,
\label{S1-0}
\end{eqnarray}
\end{subequations}

\subsubsection*{Equations of motion (\ref{A1})}
\begin{subequations}
\label{A1-1}
\begin{eqnarray}
(b_1^+)^{s-1}
&&
-m_1|S_{2,s-2}\rangle
=
|A_{1,s-1}'\rangle
,
\\
(b_1^+)^{s-2}
&&
-m_1|S_{2,s-3}\rangle
=
|A_{1,s-2}'\rangle
+l_1^+|S_{2,s-2}\rangle
,
\\
(b_1^+)^k
&&
-m_1|S_{2,k-1}\rangle
=
|A_{1k}'\rangle
+l_1^+|S_{2k}\rangle
+2r(\beta-2)l_2^+|S_{4k}\rangle
,
\\
(b_1^+)^0
&&
l_1|\Phi\rangle
=l_1^+|S_{20}\rangle
+|A_{10}'\rangle
+2r(\beta-2)l_2^+|S_{40}\rangle
,
\label{A1-0}
\end{eqnarray}
\end{subequations}

\subsubsection*{Equations of motion (\ref{A2})}
\begin{subequations}
\label{A2-1}
\begin{eqnarray}
(b_1^+)^{s-2}
&&
m_1|S_{4,s-3}\rangle
=
|S_{2,s-2}\rangle
,
\\
(b_1^+)^{s-3}
&&
m_1|S_{4,s-4}\rangle
=
|S_{2,s-3}\rangle
-l_1^+|S_{4,s-3}\rangle
,
\\
(b_1^+)^k
&&
m_1|S_{4,k-1}\rangle
=
|S_{2k}\rangle
-l_1^+|S_{4k}\rangle
,
\\
(b_1^+)^0
&&
l_2|\Phi\rangle
=
l_1^+|S_{40}\rangle
-|S_{20}\rangle
,
\label{A2-0}
\end{eqnarray}
\end{subequations}

\subsubsection*{Equations of motion (\ref{A1})}
\begin{eqnarray}
\label{A1-2}
b_2^+(b_1^+)^k
&&
|A_{1k}''\rangle=2r(\beta-2)|S_{4k}\rangle,
\end{eqnarray}

\subsubsection*{Equations of motion (\ref{S1})}
\begin{subequations}
\label{S1-2}
\begin{eqnarray}
b_2^+(b_1^+)^{s-2}
&&
|A_{2,s-2}\rangle
=
\beta r(2-\xi)|S_{2,s-2}\rangle
-m_1|A_{1,s-3}''\rangle
,
\\
b_2^+(b_1^+)^{s-3}
&&
|A_{2,s-3}\rangle
=
\beta r(2-\xi)|S_{2,s-3}\rangle
-m_1|A_{1,s-4}''\rangle
-l_1^+|A_{1,s-3}''\rangle
,
\\
b_2^+(b_1^+)^{k}
&&
|A_{2k}\rangle
=
\beta r(2-\xi)|S_{2k}\rangle
-m_1|A_{1,k-1}''\rangle
-l_1^+|A_{1k}''\rangle
,
\\
b_2^+(b_1^+)^{0}
&&
|A_{20}\rangle
=
\beta r(2-\xi)|S_{20}\rangle
-l_1^+|A_{10}''\rangle
,
\end{eqnarray}
\end{subequations}

One can show that the solution of equations of motion
(\ref{S1-1})--(\ref{S1-2}) is
\begin{eqnarray}
|S_2\rangle=|S_4\rangle=|A_1\rangle=|A_2\rangle=0.
\label{S24A12}
\end{eqnarray}
To see this one should start from the two first equations of
(\ref{S1-1}). They give
$|A_{1,s-1}'\rangle=|A_{1,s-2}'\rangle=0$.
Then we go down to next set of equations (\ref{A1-1}).
The first two of them give us
$|S_{2,s-2}\rangle=|S_{2,s-3}\rangle=0$.
Going down to the subsequent sets of equations
(\ref{A2-1}), (\ref{A1-2}), (\ref{S1-2})
we obtain one after another that
$|S_{4,s-3}\rangle=|S_{4,s-4}\rangle=0$,
$|A_{1,s-3}''\rangle=|A_{1,s-4}''\rangle=0$,
$|A_{2,s-2}\rangle=|A_{2,s-3}\rangle=0$.
After this we return to the first set of equations (\ref{S1-1})
and repeat the procedure until we obtain (\ref{S24A12}).
After this we get from (\ref{S1-0}), (\ref{A1-0}), (\ref{A2-0})
that the equations on the basic fields are (\ref{constr}).

Thus now we have
\begin{eqnarray}
|S_1\rangle=|\Phi\rangle,
&\qquad&
|S_2\rangle=|S_3\rangle=|S_4\rangle=|S_5\rangle=|A_1\rangle=|A_2\rangle=0.
\end{eqnarray}
Substituting these solutions into (\ref{A3}) we get
\begin{eqnarray}
L_2^+|S_6\rangle=0
&\Longrightarrow&
|S_6\rangle=0,
\end{eqnarray}
then substituting into (\ref{A4}) we obtain
\begin{eqnarray}
|A_4\rangle=0,
\end{eqnarray}
and finally substituting into (\ref{S5}) ones have
\begin{eqnarray}
|A_3\rangle=0.
\end{eqnarray}

Thus we remove all the auxiliary fields and the equations of
motion for the basic fields $|\Phi\rangle$
are (\ref{constr}).


\begin{thebibliography}{9}



\bibitem{reviews}
M. Vasiliev,
Higher Spin Gauge Theories in Various Dimensions,
Fortsch.Phys. 52 (2004) 702-717;
hep-th/0401177;
D. Sorokin,
Introduction to the Classical Theory of Higher Spins,
hep-th/0405069;
N. Bouatta, G. Comp\`ere and A. Sagnotti,
An Introduction to Free Higher-Spin Fields,
hep-th/0409068;
A. Sagnotti, E. Sezgin, P. Sundell,
On higher spin with a strong $Sp(2)$ conditions,
hep-th/0501156;
X. Bekaert, S. Cnockaert, C. Iazeolla, M.A. Vasiliev,
Nonlinear higher spin theories in various dimensions,
hep-th/0503128.











\bibitem{We}
I.L. Buchbinder, V.A. Krykhtin, V.D. Pershin,
On Consistent Equations for Massive Spin-2 Field Coupled to
Gravity in String Theory,
Phys.Lett. B466 (1999) 216-226,
hep-th/9908028;
I.L. Buchbinder, D.M. Gitman, V.A. Krykhtin, V.D. Pershin,
Equations of Motion for Massive Spin 2 Field Coupled to Gravity,
Nucl.Phys. B584 (2000) 615-640,
hep-th/9910188;
I.L. Buchbinder, D.M. Gitman, V.D. Pershin,
Causality of Massive Spin 2 Field in External Gravity,
Phys.Lett. B492 (2000) 161-170,
hep-th/0006144;

\bibitem{Deser}
S. Deser, A. Waldron,
Gauge Invariances and Phases of Massive Higher Spins in (A)dS,
Phys. Rev. Lett. 87 (2001) 031601,
hep-th/0102166;
Partial Masslessness of Higher Spins in (A)dS,
Nucl.Phys. B607 (2001) 577-604,
hep-th/0103198;
Null Propagation of Partially Massless Higher Spins in (A)dS and Cosmological Constant Speculations,
Phys. Lett. B513 (2001) 137-141,
hep-th/0105181;
K. Hallowell, A. Waldron,
Constant Curvature Algebras and Higher Spin Action Generating Functions,
Nucl. Phys. B724 (2005) 453-486,
hep-th/0505255.

\bibitem{Metsaev-m}
R.R.Metsaev,
Massive totally symmetric fields in AdS(d),
Phys.Lett. B590 (2004) 95-104,
hep-th/0312297.

\bibitem{Metsaev-2m}
R.R.Metsaev,
Mixed-symmetry massive fields in AdS(5),
Class.Quant.Grav. 22 (2005) 2777-2796,
hep-th/0412311;
Cubic interaction vertices of massive and massless higher spin
fields,
hep-th/0512342;

\bibitem{Zinoviev}
Yu. M. Zinoviev, On Massive High Spin Particles in (A)dS,
hep-th/0108192; Yu. M. Zinoviev, On Massive Mixed Symmetry Tensor
Fields in Minkowski Space and (A)dS, hep-th/0211233;



\bibitem{Klishevich}
S.M. Klishevich, Massive fields with arbitrary integer spin in
symmetrical Einstein space, hep-th/9812005; S.M. Klishevich, On
electromagnetic interaction of massive spin-2 particle,
hep-th/9708150; Massive fields with arbitray half-integer spin in
constant electromagnetic field, hep-th/9811030; Massive fields with
arbitray integer spin in homogeneous electromagnetic field,
hep-th/9910228; Interaction of massive integer-spin fields,
hep-th/0002024.




\bibitem{Heslop}
N. Beisert, M. Bianchi, J.F. Morales, H. Samtleben, Higher spin
symmetries and ${\cal N}=4$ SYM, JHEP 0407 (2004) 058,
hep-th/0405057; A.C. Petkou, Holography, duality and higher spin
fields, hep-th/0410116; M. Bianchi, P.J. Heslop, F. Riccioni, More
on La Grande Bouffe: towards higher spin symmetry breaking in AdS,
JHEP 0508 (2005) 088, hep-th/0504156; Paul J. Heslop, Fabio
Riccioni, On the fermionic Grande Bouffe: more on higher spin
symmetry breaking in AdS/CFT, JHEP 0510 (2005) 060, hep-th/0508086;
M. Bianchi, V. Didenko, Massive higher spin multiplets and
holography, hep-th/0502220.

\bibitem{Gates}
I.L. Buchbinder, S.J. Gates, W.D. Linch, J. Phillips, New 4d, N=1
superfiled theory: model of free massive superspin-3/2 multiplet,
Phys.Lett. B 535 (2002) 280-288, hep-th/0108200; I.L. Buchbinder,
S.J. Gates, W.D. Linch, J. Phillips, Dynamical superfiled theory of
free massive superspin-1 multiplet, Phys.Lett. B 549 (2002) 229-236,
hep-th/0207243; I.L. Buchbinder, S.J. Gates, S.M. Kuzenko, J.
Phillips, Massive 4D, N=1 superspin 1 and 3/2 multiplets and their
dualities, JHEP 0502 (2005) 056, hep-th/0501199; S. Fedoruk, J.
Lukiersky, Massive relativistic models with bosonic counterpart of
supersymmetry, Phys.Lett. B632 (2006) 371, hep-th/0506086.






\bibitem{Vasiliev}
L.Brink, R.R.Metsaev, M.A.Vasiliev,
How massless are massless fields in $AdS_d$,
Nucl. Phys. B586 (2000) 183-205,
hep-th/0005136;
K.B. Alkalaev, M.A. Vasiliev,
N=1 Supersymmetric Theory of Higher Spin Gauge Fields in
$AdS(5)$ at the Cubic Level,
Nucl.Phys. B655 (2003) 57-92,
hep-th/0206068;
K.B. Alkalaev, O.V. Shaynkman, M.A. Vasiliev,
On the Frame-Like Formulation of Mixed-Symmetry Massless Fields
in $(A)dS(d)$,
Nucl.Phys. B692 (2004) 363-393,
hep-th/0311164;
K.B. Alkalaev,
Two-column higher spin massless fields in $AdS(d)$,
hep-th/0311212;
O.V. Shaynkman, I.Yu. Tipunin, M.A. Vasiliev,
Unfolded form of conformal equations in M dimensions and
$o(M+2)$-modules,
hep-th/0401086;
E.D. Skvortsov, M.A. Vasiliev,
Geometric Formulation for Partially Massless Fields,
hep-th/0601095;
K.B. Alkalaev, O.V. Shaynkman, M.A. Vasiliev,
Frame-like formulation for free mixed-symmetry bosonic massless
higher-spin fields in AdS(d),
hep-th/0601225.

\bibitem{Metsaev-0}
R.R.Metsaev,
Free totally (anti)symmetric massless fermionic fields in
d-dimensional anti-de Sitter space,
Class.Quant.Grav. 14 (1997) L115-L121,
hep-th/9707066;
Fermionic fields in the d-dimensional anti-de Sitter spacetime,
Phys.Lett. B419 (1998) 49-56,
hep-th/9802097;
Arbitrary spin massless bosonic fields in d-dimensional
anti-de~Sitter space,
hep-th/9810231;
Light-cone form of field dynamics in anti-de Sitter space-time
and AdS/CFT correspondence,
Nucl. Phys. B563 (1999) 295-348,
hep-th/9906217;
Massless arbitrary spin fields in AdS(5)
Phys. Lett. B531 (2002) 152-160,
hep-th/0201226.


\bibitem{Sezgin}
E. Sezgin, P. Sundell,
Analysis of higher spin field equations in four-dimensions.
JHEP 0207 (2002) 055,
hep-th/0205132;
Holography in 4D (Super) Higher Spin Theories and a Test via Cubic Scalar Couplings,
JHEP 0507 (2005) 044,
hep-th/0305040;
An Exact Solution of 4D Higher-Spin Gauge Theory,
hep-th/0508158.


\bibitem{Sagnotti}
D. Francia, A. Sagnotti,
Free geometric equations for higher spins,
Phys. Lett. B543 (2002) 303-310,
hep-th/0207002;
On the geometry of higher-spin gauge fields,
Class. Quant. Grav. 20 (2003) S473-S486,
hep-th/0212185;
Minimal Local Lagrangians for Higher-Spin Geometry,
Phys. Lett. B624 (2005) 93-104,
hep-th/0507144;
A.  Sagnotti, M.  Tsulaia, On higher spins and the tensionless limit of
String Theory, Nucl. Phys.  B682 (2004) 83-116, hep-th/0311257;
 A. Fotopoulos, K.L. Panigrahi, M. Tsulaia,
On Lagrangian formulation of Higher Spin Theories on AdS,
hep-th/0607248.


\bibitem{Rajan}
F. Kristiansson, P. Rajan,
Scalar Field Corrections to AdS$_4$ Gravity from Higher Spin Gauge Theory,
JHEP 0304 (2003) 009,
hep-th/0303202.

\bibitem{Bekaert} 
X.~Bekaert, N.~Boulanger and S.~Cnockaert, 
Spin three gauge theory revisited, 
JHEP 0601 (2006) 052,
hep-th/0508048;
N.~Boulanger, S.~Leclercq and S.~Cnockaert,
Parity violating vertices for spin-3 gauge fields,
Phys. Rev. D73 (2006) 065019,
hep-th/0509118.



\bibitem{Bonelli}
G. Bonelli, On the Tensionless Limit of Bosonic Strings, Infinite
Symmetries and Higher Spins, Nucl. Phys. B669 (2003) 159-172,
hep-th/0305155; G. Barnich, G. Bonelli, M. Grigoriev, From BRST to
light-cone description of higher spin gauge fields, hep-th/0502232.


\bibitem{Sorokin}
M. Plyushchay, D. Sorokin and M. Tsulaia, Higher Spins from
Tensorial Charges and $OSp(N|2n)$ Symmetry, JHEP 0304 (2003) 013,
hep-th/0301067; GL Flatness of $OSp(1|2n)$ and Higher Spin Field
Theory from Dynamics in Tensorial Space, hep-th/0310297; I. Bandos,
X. Bekaert, J.A. Azcarraga, D. Sorokin, M. Tsulaia, Dynamics of
higher spin fields and tensorial space, JHEP 0505 (2005) 031,
hep-th/0501113; E. Ivanov, J. Lukiersky, Higher spins from nonliner
realizations of $OSp(1|8)$, Phys.Lett. B624 (2005) 304,
hep-th/0505216; S. Fedoruk, E. Ivanov, Master higher spin particle,
hep-th/0604111; S. Fedoruk, E. Ivanov, J. Lukiersky, Massless higher
spin D=4 superparticle with both ${\cal N}=1$ supersymmetry and its
bosonic counterpart, hep-th/0606053.


\bibitem{Barnich}
G.
Barnich, M.  Grigoriev, A.  Semikhatov, I.  Tipunin, Parent field
theory and unfolding in BRST first-quantized terms, Commun. Math. Phys.
260 (2005) 147-181, hep-th/0406192; Glenn Barnich, Maxim Grigoriev,
Parent form for higher spin fields on anti-de Sitter space,
hep-th/0602166.

\bibitem{BFV}E.S. Fradkin, G.A. Vilkovisky, Quantization of
relativistic system with constraints, Phys.Lett. B55 (1975) 224;
I.A. Batalin, G.A. Vilkovisky, Relativisitic S-matrix of dynamical
systems with bosonic and fermionic constraints, Phys.Lett. B69
(1977) 309; I.A. Batalin, E.S. Fradkin, Operator quantization of
relativistic dynamical systems subject to first class constraints,
Phys.Lett. B128 (1983) 303.

\bibitem{bf}
I.A. Batalin, E.S. Fradkin, Operator quantizatin method and
abelization of dynamical systems subject to first class constraints,
Riv. Nuovo Cimento, 9, No~10 (1986) 1; I.A. Batalin, E.S. Fradkin, Operator
quantization of dynamical systems subject to constraints.  A further
study of the construction, Ann. Inst. H. Poincare, A49 (1988) 145; M.
Henneaux, C. Teitelboim, Quantization of Gauge Systems, Princeton
Univ. Press, 1992.

\bibitem{PF}M. Fierz, W. Pauli, On relativistic wave equations for
particles of arbirary spin in an electromagnetic field,
Proc.R.Soc.London, Ser. A, 173 (1939) 211-232.







\bibitem{Fronsdal}
C. Fronsdal,
Singletons and massless, integral-spin fields on de Sitter space,
Phys. Rev. D20 (1979) 848--856.




\bibitem{BRST-AdS}
I.L. Buchbinder, A. Pashnev and M Tsulaia,
Lagrangian formulation of the massless higher integer spin
fields in the AdS background,
Phys. Lett. B523 (2001) 338-346,
hep-th/0109067;
Massless Higher Spin Fields in the AdS Background and BRST
Constructions for Nonlinear Algebras,
hep-th/0206026;
X. Bekaert, I.L. Buchbinder, A. Pashnev, M. Tsulaia,
On Higher Spin Theory: Strings, BRST, Dimensional Reductions,
Class.Quant.Grav. 21 (2004) S1457-1464, hep-th/0312252.

\bibitem{0206027}
C. Burdik, O. Navratil, A. Pashnev,
On the Fock Space Realizations of Nonlinear Algebras Describing the High Spin Fields in AdS Spaces,
hep-th/0206027.




\bibitem{0101201}
C. Burdik, A. Pashnev, M. Tsulaia, On the mixed symmetry irreducible
representations of the Poincare group in the BRST approach,
Mod.Phys.Lett. A16 (2001) 731--746, hep-th/0101201.


\bibitem{Sevrin}
K. Schoutens, A.Sevrin and P. van Nieuwenhuizen,
Quantum BRST Charge for Quadratically Nonlinear Lie Algebras,
Commun. Math. Phys. 124, 87--103 (1989).


\bibitem{W}E. Witten, Noncommutative geometry and string field
theory, Nucl.Phys. B268 (1986) 253.

\bibitem{SFT}C.B. Torn, String field thoory, Phys.Repts 175 (1989)
1-101; W. Taylor, B. Zwiebach, D-branes, tachyons and string field
theory, hep-th/0311017.


\bibitem{Ouvry}S. Ouvry, J. Stern, Gauge fields of any spin and
symmetry, Phys.Lett. B177 91986) 335-340; A.K.H. Bengtsson, A
unified action for higher spin gauge bosons from covariant string
theory, Phys.Lett. B182 91986) 321-325; W. Siegel, B. Zwiebach,
Gauge string fields from light cone, Nucl.Phys. B282 (1987) 125; W.
Siegel, Gauging Ramond string fields via $OSp(1,1|2)$, Nucl.Phys.
B284 (1987) 632.


\bibitem{0410215}
I.L. Buchbinder, V.A. Krykhtin, A. Pashnev,
BRST approach to Lagrangian construction for fermionic massless
higher spin fields,
Nucl. Phys. B711 (2005) 367--391,
hep-th/0410215.

\bibitem{0505092}
I.L. Buchbinder, V.A. Krykhtin,
Gauge invariant Lagrangian construction for massive bosonic
higher spin fields in D dimensions,
Nucl. Phys. B727 (2005) 536--563,
hep-th/0505092.



\bibitem{0511276}
I.L. Buchbinder, V.A. Krykhtin,
BRST approach to higher spin field theories,
hep-th/0511276.




\bibitem{0603212}
I.L. Buchbinder, V.A. Krykhtin, L.L. Ryskina, H. Takata,
Gauge invariant Lagrangian construction for massive higher spin
fermionic fields,
hep-th/0603212.

\bibitem{Beng}I.G. Koh, S. Ouvry, Interacting gauge fields of any spin and symmetry,
Phys.Lett. B179 (1986) 115;
A.K.H. Bengtsson, BRST approach to interacting higher
spin fields, Class.Quant.Grav. 5 (1988) 437; L. Cappiello, M.
Knecht, S. Ouvry, J. Stern, BRST construction of interacting gauge
theories of higher spin fields, Ann.Phys. 193 (1989) 10-39; F.
Fougere, M. Knecht, J. Stern, Algebraic construction of higher spin
interaction vertices, preprint LAPP-TH-338/91.

\bibitem{conversion}L.D. Faddeev, S.L. Shatoshvili, Realization of
the Schwinger term in the Gauss low and the possibility of correct
quantization of a theory with anomalies, Phys.Lett. B167 (1986)
225-338; I.A. Batalin, I.V. Tyutin, Existence theorerm for the
effective gauge algebra in the generalized canonical formalism and
Abelian conversion of second class constraints, Int.J.Mod.Phys. A6
(1991) 3255-3282; E. Egorian, R. Manvelyan, Quantization of
dynamical systems with first and second class constraints,
Theor.Math.Phys. 94 (1993) 241-252.


\end{thebibliography}
\end{document}